\documentclass{article}

\usepackage{arxiv}

\usepackage{algorithm}
\usepackage{algorithmic}
\usepackage{amsfonts}
\usepackage{amsmath}
\usepackage{amssymb}
\usepackage{array}
\usepackage{caption}
\usepackage{cite}
\usepackage{color}
\usepackage{comment}
\usepackage{enumerate}
\usepackage{epstopdf}
\usepackage{graphicx}
\usepackage{hyperref}       
\usepackage{float}
\usepackage{multicol}  
\usepackage{multirow}
\usepackage{soul}
\usepackage{subcaption}
\usepackage{tkz-graph}
\usepackage{url}

\begin{document}

\title{Learning Binary Color Filter Arrays with Trainable Hard Thresholding}

\author{
    \href{https://orcid.org/0000-0002-3743-4000}{\includegraphics[scale=0.06]{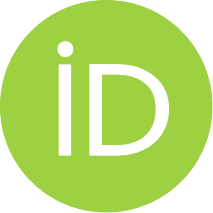}\hspace{1mm}Cemre O. Ayna} \\
	Department of Electrical \& Computer Engineering\\
	Mississippi State University\\
	Mississippi State, MS, 39759 \\
	\texttt{ca1389@msstate.edu} \\
	\and
    \href{https://orcid.org/0000-0003-0779-9620}{\includegraphics[scale=0.06]{orcid.pdf}\hspace{1mm}Bahadir K. Gunturk} \\
	Department of Electrical-Electronics Engineering\\
	Medipol University\\
	Istanbul, Turkey, 34810 \\
	\texttt{bkgunturk@medipol.edu.tr} 
    \and
    \href{https://orcid.org/0000-0001-8923-0299}{\includegraphics[scale=0.06]{orcid.pdf}\hspace{1mm}Ali C. Gurbuz} \\
	Department of Electrical \& Computer Engineering\\
	Mississippi State University\\
	Mississippi State, MS, 39759 \\
	\texttt{gurbuz@ece.msstate.edu} 
\thanks{
This work was supported by the National Science Foundation under Grant No. 2047771 (Corresponding author: Ali C. Gurbuz.)
Dr. Ali Cafer Gurbuz and  Cemre Omer Ayna are with the Department of Electrical and Computer Engineering at Mississippi State University, MS-39762, US. (email: gurbuz@ece.msstate.edu). Dr. Bahadir Gunturk is with the Istanbul Medipol University (email: bkgunturk@medipol.edu.tr). } 
}

\maketitle

\begin{abstract}
Color Filter Arrays (CFA) are optical filters in digital cameras that capture specific color channels. Current commercial CFAs are hand-crafted patterns with different physical and application-specific considerations. This study proposes a binary CFA learning module based on hard thresholding with a deep learning-based demosaicing network in a joint architecture. Unlike most existing learnable CFAs that learn a channel from the whole color spectrum or linearly combine available digital colors, this method learns a binary channel selection, resulting in CFAs that are practical and physically implementable to digital cameras. The binary selection is based on adapting the hard thresholding operation into neural networks via a straight-through estimator, and therefore it is named HardMax. This paper includes the background on the CFA design problem, the description of the HardMax method, and the performance evaluation results. The evaluation of the proposed method includes tests for different demosaicing models, color configurations, filter sizes, and a comparison with existing methods in various reconstruction metrics. The proposed approach is tested with Kodak and BSDS500 datasets and provides higher reconstruction performance than hand-crafted or alternative learned binary filters. 
\end{abstract}


\section{Introduction} \label{sec:introduction}
A digital camera captures an image by exposing its sensor array in which each sensor corresponds to one pixel in the final image to the incoming light for a certain amount of time. A Color Filter Array (CFA) is an optical filter placed on a camera sensor array. Sensors alone cannot differentiate between individual colors; instead, CFAs facilitate capturing color information by sifting only one frequency band in the visible light spectrum corresponding to the selected color per pixel. The raw input of the filtered camera sensor array corresponds to an image in which each pixel contains the intensity information of only one color channel and lacks the rest. 

\begin{figure}[!ht]
    \centering
    \begin{subfigure}{0.2\textwidth}
        \centering
        \includegraphics[width=0.8\textwidth]{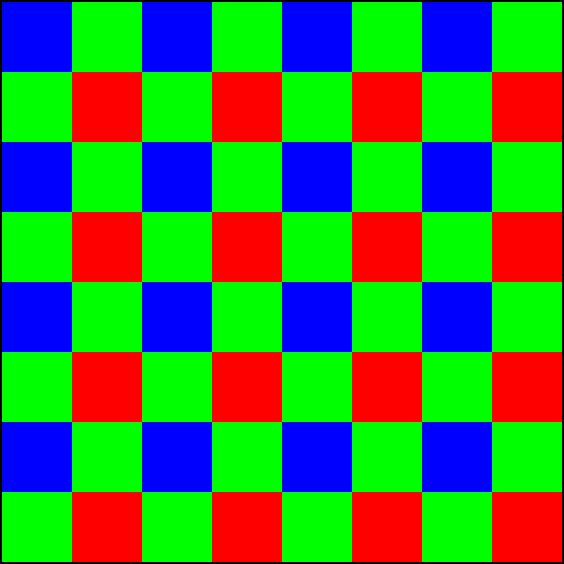}
        \caption{}
    \end{subfigure}
    \begin{subfigure}{0.2\textwidth}
        \centering
        \includegraphics[width=0.8\textwidth]{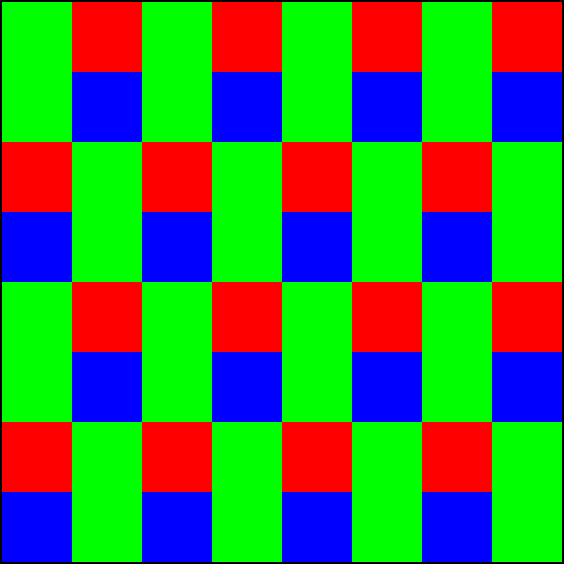}
        \caption{}
    \end{subfigure}
    \begin{subfigure}{0.2\textwidth}
        \centering
        \includegraphics[width=0.8\textwidth]{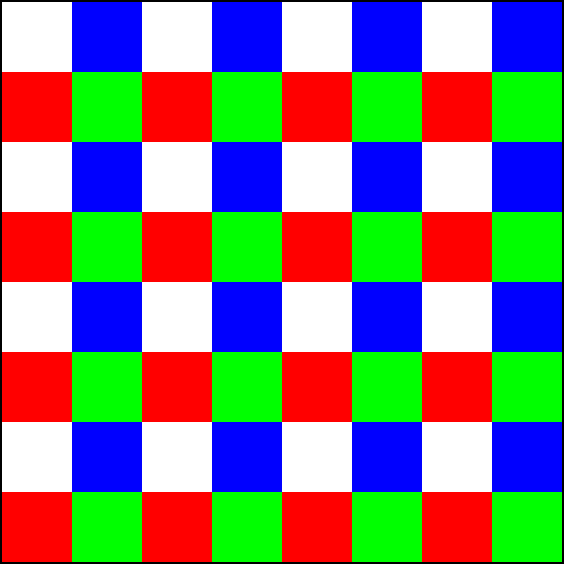}
        \caption{}
    \end{subfigure}
    \begin{subfigure}{0.2\textwidth}
        \centering
        \includegraphics[width=0.8\textwidth]{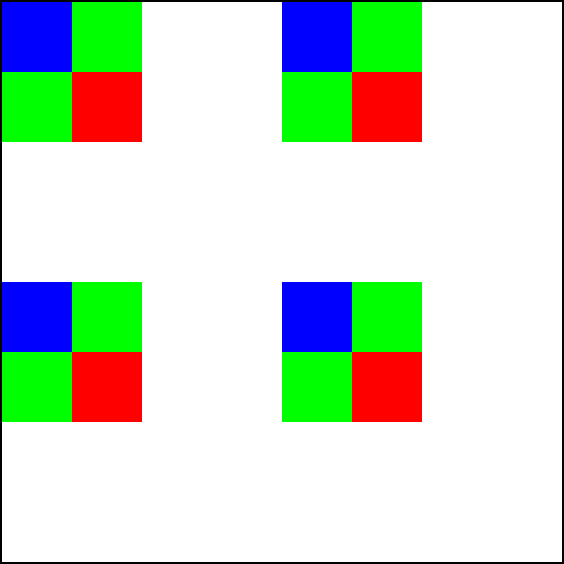}
        \caption{}
    \end{subfigure}
    \caption{The fixed hand-crafted CFA examples used in this study for evaluation: (a) Bayer, (b) Lukac, (c) RGBW, and (d) CFZ filters. Each pattern is extended to $8 \times 8$ size for visualization.}
    \label{fig:cfa_examples}
\end{figure}

Virtually all available commercial CFAs are designed by hand with different considerations depending on the camera and environment characteristics. The most commonly used CFA pattern is the Bayer filter \cite{bayer1976color}. Several other hand-crafted CFAs are also present in specific camera models such as Lukac filter \cite{lukac2005color}, Kodak’s CYYM filter, Fujifilm’s X-Trans \cite{tanaka2013xtrans}, CMWY filter, and Compton’s RGBW filter along with Kodak’s RGBW filter variations \cite{chung2020effective}. The hand-crafted filters used in this study for evaluation are illustrated in Figure~\ref{fig:cfa_examples}. The study \cite{lukac2005color} provides an extensive review and analysis of the importance of CFA design on the final image.

The process following the color filtering operation in digital image acquisition is known as demosaicing, which is estimating the unknown color values in raw camera returns \cite{gunturk2005demosaicking}. Common demosaicing algorithms are mainly spatial or frequency domain interpolation-based techniques \cite{malvar2004high, lukac2004novel}. These algorithms show variation for specific applications and different CFA types \cite{kimmel1999demosaicing, li2005demosaicing, lukac2006new}. Detailed reviews of the classical demosaicing algorithms for various CFAs are available in \cite{li2008image, safna2018colour}. A unique CFA design requires a dedicated demosaicing algorithm depending on the filter pattern and available information from the captured scene. 

With the emergence of machine learning (ML) and neural networks (NN) in computational imaging, various studies that suggest using NNs in demosaicing or joint demosaicing-denoising pipelines \cite{gharbi2016deep, de2021data, park2019color, tan2018deepdemosaicking, kokkinos2019iterative} have appeared. These approaches work with raw camera return and propose various NN architectures for mapping the inputs to the full-color image. These methods show that enhanced reconstruction quality and computational speed can be achieved with deep learning (DL).

Recent studies applied ML solutions for learning a CFA pattern to address the issue of exploiting the features of natural images for high-quality full-color image reconstruction \cite{tang2021demosaicing, henz2018deep}. Although working in the RGB domain, these solutions learn CFAs that employ the full digital color spectrum; their process learns a linear combination of all three color channels. This approach results in learning a unique color per sensor. Although weighted combinations of colors provide enhanced reconstructions compared to fixed CFAs like Bayer, these learned filters are impractical for physical implementation in commercial cameras where each pixel reads a single color from the digital color configuration. Some other studies assume working in the multispectral domain and learning MultiSpectral Filter Arrays (MSFA) \cite{bian2021generalized, zhang2021jointly}. Although there is active research for building commercial cameras with MSFAs, these cameras are still in the prototype phase due to their high production cost and the raw image formats requiring additional operations to get the usual RGB images as the final product. Section~\ref{subsec:cfa_learning} includes a more detailed discussion on the applicability issues of MSFAs. 

Modern commercial digital cameras use color configurations with a few colors, with most of them opting for RGB configuration. For this reason, it is necessary to develop learned binary CFAs that utilize only one color channel at each pixel and still provide enhanced image reconstruction compared to hand-crafted CFAs. To the best of our knowledge, only one method presents a way to learn a true binary CFA pattern in RGBW configuration \cite{chakrabarti2016learning}. This study adapts SoftMax operation with a scalar value that increases in time exponentially in order to acquire quasi-binary filter weights from output weights.

This paper proposes an alternative method for learning a binary CFA in a joint filter-demosaicer architecture. The proposed joint framework is an end-to-end architecture with two modules; the head module learns a constrained binary CFA during training, and the tail module reconstructs a color image from filtered raw camera images. This joint learning approach enforces the learned binary CFA to be optimal for color image demosaicing. The proposed method adapts hard thresholding as band selection operation in the CFA learning module that is compatible with stochastic gradient descent. The CFAs learned with this method can be used in camera sensor arrays without the impracticality concerns since the selection of only one color channel is enforced for each pixel. Our results indicate CFAs learned with this approach provide a higher reconstruction performance than the hand-designed filters and the alternative proposed in \cite{chakrabarti2016learning}. With reference to the hard thresholding as the basis of this process, we named our binary CFA learning module \emph{HardMax}. The novelties of this study can be described in the following points.

\begin{itemize}
    \item This study presents an NN model for joint color filtering - demosaicing modules with a novel binary CFA learning mechanism and an indigenous high-performing demosaicer architecture.
    \item The proposed CFA learning method (HardMax) is adaptable to different joint DL architectures, allowing us to learn optimum CFAs for different objectives of the architectures following the CFA module such as reconstruction or classification.
    \item Unlike other proposed methods, HardMax aims to find an optimum binary CFA so that the learned CFA can be easily applied to commercial digital cameras.
    \item This study includes the evaluation and analysis of different parameters that affect the learned CFAs, such as filter size, color configuration, and training data size.
\end{itemize}

The rest of the paper follows this structure; Section \ref{sec:background} gives a background on the available literature on the machine learning-based CFA design. Section \ref{sec:method} describes the proposed CFA learning and demosaicing method along with the training and evaluation procedures. Section \ref{sec:setup} describes the dataset, training process, and evaluation. Section \ref{sec:results} presents results, the obtained performance metrics, and the comparison with the existing approaches. Section \ref{sec:discussion} includes discussion on the proposed approach and results, the shortcomings of the study, and the potential road map for future work. Finally, section \ref{sec:conclusion} draws the conclusions.

\section{Background} \label{sec:background}

\subsection{Machine Learning in CFA Design} \label{subsec:cfa_learning}

\begin{figure}[!ht] 
    \captionsetup[subfigure]{justification=centering}
    \centering
    \begin{subfigure}{0.33\textwidth}
        \centering
        \includegraphics[width=0.8\textwidth]{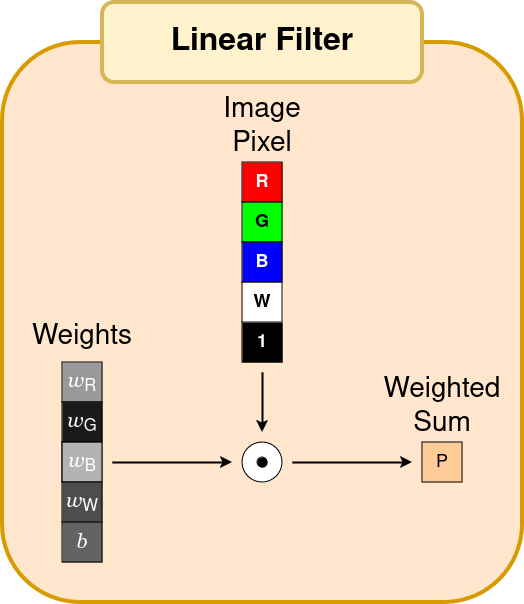}
        \caption{}
        \label{fig:linear}
    \end{subfigure}
    \hspace{12mm}
    \begin{subfigure}{0.36\textwidth}
        \centering
        \includegraphics[width=1\textwidth]{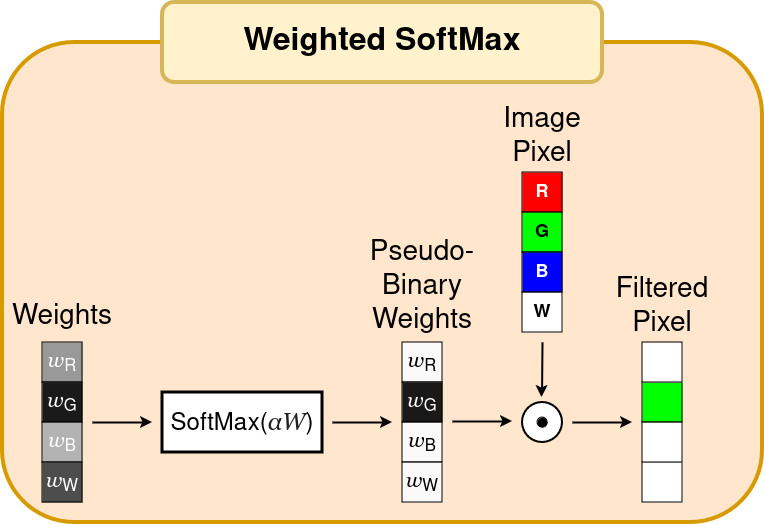}
        \caption{}
        \label{fig:weighted_softmax}
    \end{subfigure}
    \caption{Visualization of the CFA learning methods used in this study for comparison. (a) Linear \cite{henz2018deep}, (b) Weighted SoftMax \cite{chakrabarti2016learning}. Figures show the RGBW configuration.}  
    \label{fig:alt_cfa_learning}
\end{figure}

Compared to demosaicing, the volume of studies on the ML-based CFA design problem is small, with the recent literature focusing on MSFAs. Some of these methods are designed for hyperspectral image acquisition \cite{jacome2021deep, wu2019optimized, sawyer2022opti}, while others target commercial digital cameras \cite{bian2021generalized, zhang2021jointly, li2023jointly}. However, multispectral image acquisition has several issues that prevent it from being used in commercial digital cameras. The first problem is the high cost of multispectral cameras due to the complexity of their production. The second issue is that a higher number of frequency bands causes the curse of dimensionality for a full-color image. Thirdly, the larger interval of missing channel values corresponds to lower resolution in the final product and more complex demosaicing algorithms. For these reasons, this study focuses on solutions for CFAs that use the available number of colors in digital image (usually three chromatic and one luminance “white” channel).

The lack of available literature on ML-based CFA design is even more striking when the literature search is constrained to the RGB case. There is only a handful of studies for modern ML techniques in CFA design \cite{henz2018deep, li2017optimized}. As an example, Henz et al. \cite{henz2018deep} introduced an autoencoder architecture as a joint filtering-demosaicing pipeline where the encoder is composed of a 3D tensor with weights for each color channel (three nonnegative unconstrained weight sets for each color channel) plus an unconstrained bias term used in a matrix multiplication with the full-color image in the training process (see Figure~\ref{fig:linear}). This method is included in the evaluation and will be called Linear filter in the rest of the paper for convention. Li et al. \cite{li2017optimized} propose a CFA design algorithm based on representing CFAs as overcomplete dictionaries that sample an original image and finding the set of dictionary values that minimize the mutual coherence value, for mutual coherence is an important factor in signal reconstruction and smaller coherence corresponds to higher quality reconstruction. In practice, Li's filter corresponds to a similar linear projection operation as in \cite{henz2018deep} without a bias term. 

Even though the studies \cite{henz2018deep, li2017optimized} define their search domain on the RGB channel, their final product is not a purely binary RGB filter but a filter with different linear combinations of available channels for each pixel. That is due to the mentioned studies defining their CFA learning process as a linear sum operation on the RGB spectrum. In return, the unconstrained linear combination of non-negative weights leads to a quasi-infinite number of potential colors for selection. In order to learn a CFA from available color channels, the CFA learning problem must be redefined as a constrained selection problem rather than a weighted linear sum. There is only one study known to us that adopts such a strategy; Chakrabarti's method in \cite{chakrabarti2016learning} uses a soft thresholding adaptation of binary selection by applying SoftMax on a set of weights for each color channel per pixel (Figure~\ref{fig:weighted_softmax}). The SoftMax operation is also controlled by a separate scalar value for weights before the SoftMax operation. The scalar is initialized as a small value. As training progresses, this scalar grows to binarize the elements of the output vector. This scalar value increases as training processes, stretching the SoftMax output more into the limit values of 0 and 1. For convenience, this approach is named as Weighted SoftMax in this paper.

A recent study \cite{bai4753575joint} learns a separate CFA but essentially uses the same CFA learning scheme as in \cite{chakrabarti2016learning}. Several regularization enforced binary coded aperture learning mechanisms are yet to be adapted to color filter selection \cite{arguello2022deep, bacca2021deep}. This paper aims to introduce a new approach for learning a constrained binary CFA that leads to a higher reconstruction performance than the available solution.

\subsection{DL-Based Demosaicing} \label{subsec:dl_based_demos}
There is a plethora of demosaicing algorithms designed to work with different CFA filters or address problems in image recovery. The classical approaches can be grouped into three categories according to their strategies \cite{li2008image}; complex interpolation algorithms accounting geometry or optics \cite{hirakawa2007spatio, malvar2004high}, heuristic algorithms that build the digital image upon iterative recovery (for instance, recovering luminance, then chrominance) \cite{lu2003color, kakarala2002adaptive}, or statistical models assuming an interpixel or interchannel dependency (like sparsity-based interpolation methods) \cite{li2005demosaicing, portilla2005low}. Because of the lack of information about the interpixel and interchannel dependency and the high variability of these values, each approach has its drawbacks and issues.

The DL-based demosaicing methods emerged after the neural networks proved to be efficient function approximators in many computer vision applications. It is important to note that the demosaicing problem is an undetermined reconstruction problem \cite{jin2017deep}, and this fact is exploited heavily by the sparse representation-based demosaicing algorithms that employ dictionary search or other methods developed for CS \cite{li2005demosaicing, portilla2005low}. For the same reason, the available DL-based demosaicing models share similarities with the DL-based image reconstruction and super-resolution models.

Here, we present the three DL-based demosaicing models selected for comparison during the evaluation of our demosaicing model in this study. The reason for selecting these studies is that just like our proposed framework all three models are used alongside a CFA learning algorithm in their respective studies. Common to all approaches, we assume an image patch of the size $N \times N \times C$ as the final output, where $N$ is the size of both the width and the height, and $C$ is the number of channels. This patch is one of the all non-overlapping adjacent patches extracted from the original image. 

The first demosaicing architecture is found in \cite{henz2018deep} (Linear) and it was inspired by autoencoders and fully convolutional network architectures. The input is an $N \times N \times (2C+1)$-sized feature map, which includes $C$ number of individual color channels (called submosaics), the $C$ number of interpolated submosaics with a k-neighboring kernel, and the monochromatic raw image. The separate inclusion of submosaics and their interpolated version is to help the network to skip the procedure of interpolation and channel separation in order to let the model focus on the reconstruction. The interpolations of submosaics are created and concatenated to the raw image with an interpolation kernel between the filtering (encoding) and demosaicing (decoding) operations. The rest of the demosaicing model is a fully convolutional network with 12 layers in total. The kernel size is fixed to $3 \times 3$. The first six layers have 64 kernels, while the last six have 128. At the end of the model, the input raw camera sensor matrix is concatenated with the output of the last convolutional layer. Then, this tensor is passed through a final convolutional layer to get the reconstructed full-image patch.

The second demosaicing architecture is adapted from the compared alternative study (Weighted SoftMax) \cite{chakrabarti2016learning}. In this architecture, the input is a $3N \times 3N \times 1$-size raw camera measurement of a $N \times N$ image patch with its neighboring area, and the output is a $N \times N \times C$-size reconstructed central full-color image block. The reason for using surrounding patches is to use the information around the central patch to reinforce the reconstruction quality and prevent artifacts. The demosaicing architecture includes two parallel streams creating color channel priors. The first stream consists of a fully connected (FC) layer with $P \times P \times 3K$ number of neurons followed by a reshaping operation and a $1 \times 1$ convolutional layer. The purpose of this stream is to extract all the color information from the raw sensor readings. In the original study, the FC layer is preceded by a natural logarithm and succeeded by exponential operation. In the evaluations, this approach caused the training accuracy to fluctuate and even diverge from a solution; thus, we had to discard it. The second stream is an encoder composed of a group of convolutional layers with an $F$ number of kernels and ReLU activation function, followed by an FC layer with $N \times N \times 3K$ neurons and a reshape operation. This stream’s purpose is to capture spatial features independent of the color channel to augment the estimation of the absent color values. In our comparison, $F$ and $K$ values are 128 and 32. The same values were used in this study. 

The third demosaicing architecture is based on a DL-based demosaicing model proposed in \cite{de2021data}. In the suggested design, the raw camera return is processed in two different modules to reconstruct two different information: low-frequency color (chrominance) and high-frequency shapes (luminance). The luminance reconstruction network returns a single matrix intended to learn the grayscale information which carries most of the low-level information, such as edges and patterns. This network consists of only one hidden convolutional layer and one output convolutional layer. Since there are no further details on this network, we chose the filter sizes for all layers as $3 \times 3$, while the hidden layer has 64 filters and the output layer has one filter. On the other hand, the chrominance network is an autoencoder and returns an $N \times N \times 3$ size output. Like the luminance network, the original paper does not mention the actual architecture. For this reason, this study devised an autoencoder with 3 convolutional layers and 3 deconvolutional in total. The first three layers have 64, 128, and 256 $3 \times 3$ sized kernels with $2 \times 2$ stride respectively. The three deconvolutional layers following this recover the same shape to create the $N \times N \times 3$ color information matrix. The outputs of the luminance and the chrominance networks are then summed up to create the final reconstruction. 

\section{Proposed Method} \label{sec:method}
The proposed HardMax layer for learning CFA along with the demosaicer model is an extension of the study presented in \cite{ayna2024learning}. In this section, we will explain the details of the binary CFA learning algorithm, the architecture of the proposed demosaicer model, and the use of both modules in a joint framework.

 \begin{figure*}[!ht]
    \centering
    \includegraphics[width=0.98\textwidth]{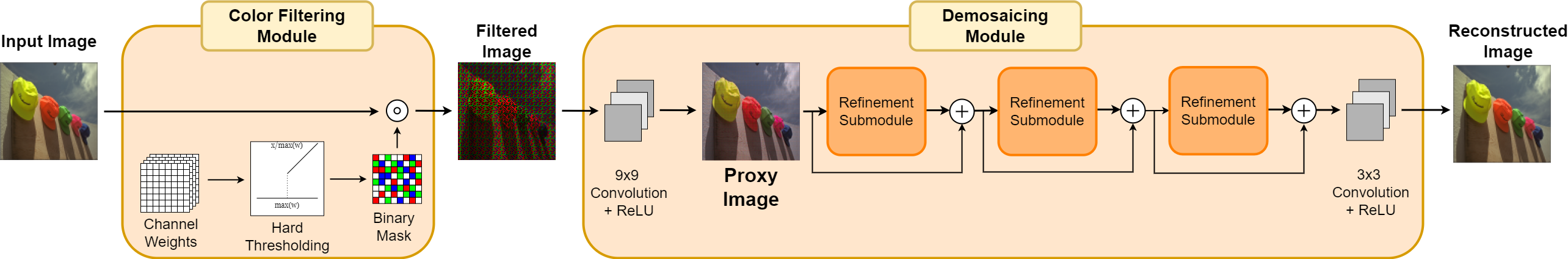}
      \caption{The full joint model in forward passing in the RGBW configuration. The output of the color filtering module represents the raw sensor input and is passed to the demosaicing module as input.}  
      \label{fig:joint_model}
\end{figure*}

\subsection{Joint Binary CFA Learning and Demosaicing Architecture} \label{subsec:joint}
An important aspect of DL-based solutions is their capability to combine multiple optimization problems into a single framework. For our framework, we define two separate objectives. The primary objective is to learn a binary CFA, and the secondary objective is to reconstruct a color image from raw sensor inputs with the given binary CFA. The goal is to achieve both these objectives together in a single DL architecture that learns binary CFAs that result in high reconstruction performance. Mainly, joint frameworks have the advantage of working in a combined search space, therefore eliminating the risk of overshooting in the overall process while performing singular tasks as well as separate models. Different versions of joint CFA and demosaicing solutions have also been used in the studies \cite{henz2018deep,chakrabarti2016learning} as detailed in Section \ref{sec:background}. 


The full neural network model proposed in this study is a combination of the binary CFA learning module detailed in Section~\ref{subsec:max_threshold} and the demosaicing model based on image reconstruction networks described in Section~\ref{subsec:demos}. Figure~\ref{fig:joint_model} shows a visual representation of the full model in a single forward propagation. During training, the full model takes an $N \times N \times C$ image patch $x$ as input and returns a reconstruction of the image patch $\hat{x}$ as output. The first module of the joint model is a binary CFA learning module that acts as a simulated color filter and returns an output $y$ which corresponds to raw camera sensor observations. This output is then passed into the demosaicing model for reconstruction of the full-color image. The pseudoimage $\hat{x}^{(0)}$ returned by the first convolution operation of the demosaicing module is refined by consecutive refinement submodules. The complete architecture is learned jointly by minimizing a common loss function enforcing mean square error between reconstructed and labeled images to be smaller. 

The main consideration in any end-to-end DL framework is guaranteeing differentiability in every step. The current method of solution searching in NNs involves stochastic gradient descent (SGD), an update algorithm that uses the gradient operation in calculating weight change increments. This is a critical problem specifically in the binary CFA learning module as it will be detailed in the next section. 

\subsection{HardMax: The Binary CFA Learning Module} \label{subsec:max_threshold}

\begin{figure}[ht!]
\captionsetup[subfigure]{justification=centering}
    \centering
    \begin{subfigure}{0.4\textwidth}
        \centering
        \includegraphics[width=0.9\textwidth]{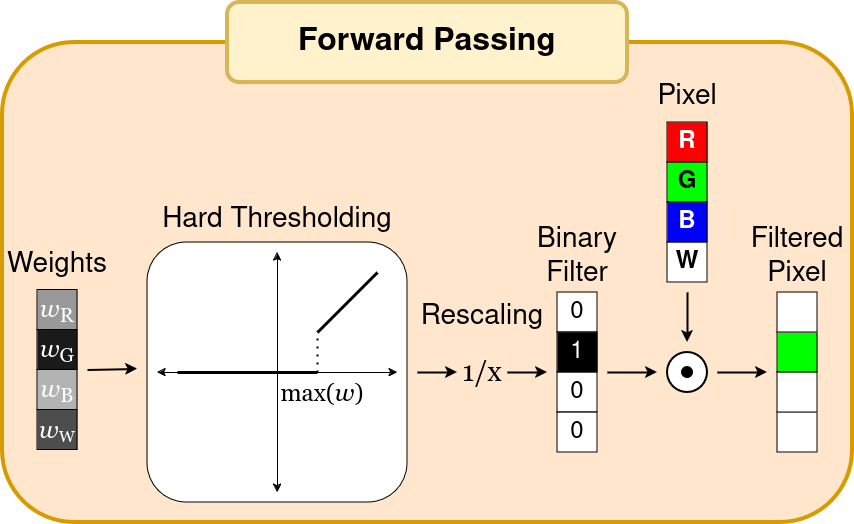}
        \caption{}
        \label{fig:max_threshold_forward}
    \end{subfigure}
    \begin{subfigure}{0.4\textwidth}
        \centering
        \includegraphics[width=0.9\textwidth]{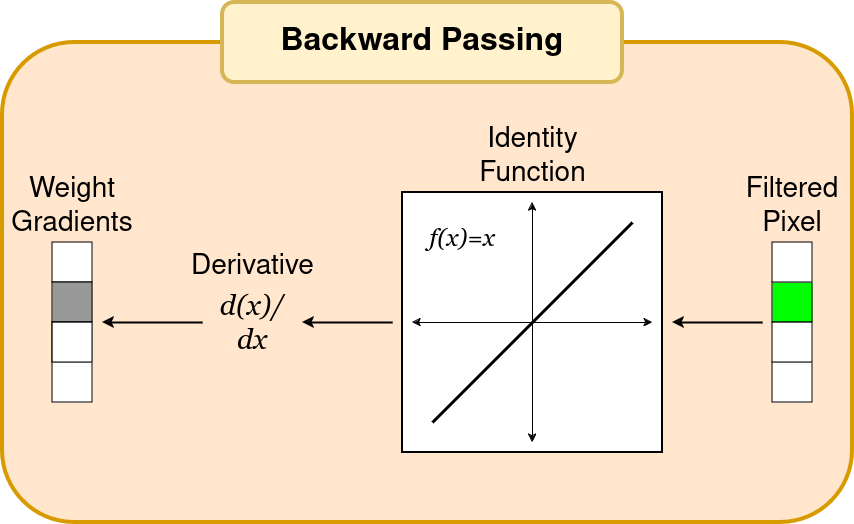}
        \caption{}
        \label{fig:max_threshold_backward}
    \end{subfigure}
    \caption{Description of the HardMax filter in (a) forward propagation phase and (b) backpropagation phase during training. The binary selection is expanded as thresholding by the maximum weight value and normalization operations in the network.}  
    \label{fig:max_threshold}
    \vspace{-2mm}
 \end{figure}

The problem of adapting the binary selection operation in neural networks originates from the thresholding function being non-differentiable. A solution to the problem can be achieved with the Straight-Through Estimator (STE). STE is gradient estimation for functions without differentials in potential parameter points \cite{bengio2013estimating}. Assigning an STE to a non-differentiable function begins with observing the function's behavior in the concerned domain, then looking for a set of potential functions which are following an output similar to the actual function and are differentiable in that domain. The basic idea of our application was based on the study in \cite{feng2020convolutional}. In the case of binary selection, we redefine the process first as hard thresholding followed by normalization. In hard thresholding, the gradient is carried only when the weight value is higher than the threshold value, and this value is equal to one. An alternative function that behaves exactly the same in the relevant domain is the identity function. Therefore, the identity function’s derivative (i.e., just the value 1) can be selected as the STE of the hard thresholding operation. 

With a solution for direct implementation of hard thresholding provided, we assume the problem of sampling an image block with $C$ channels and $N \times N$ size. The objective of the learnable discrete CFA module is to select one channel in every pixel and discard the rest, creating $N \times N$ number of measurements as an output. The HardMax module manages this by initializing a $N \times N \times C$ sized tensor with uniformly random values between the interval $[0,1]$ as real-valued weights. The weights don’t have to be constrained into a specific interval, and their sign is irrelevant since the actual information encoded with the weights is their relative greatness to each other. The highest weight value in each pixel represents the selected channel. This weight is defined as the threshold value. All pixel weights are passed into thresholding, as shown in Fig~\ref{fig:max_threshold}. The resulting vector is then binarized to convert the actual weight value into a binary mask. 

The overall masking process is simple to implement, and with a preassigned gradient value, the backpropagation operation is computationally efficient with $N^2$ operations in total. For this reason, HardMax takes small computational resources and time both in training and evaluation. 

\subsection{Demosaicing Module} \label{subsec:demos}

 \begin{figure*}[ht!]
    \centering
    \includegraphics[width=0.8\textwidth]{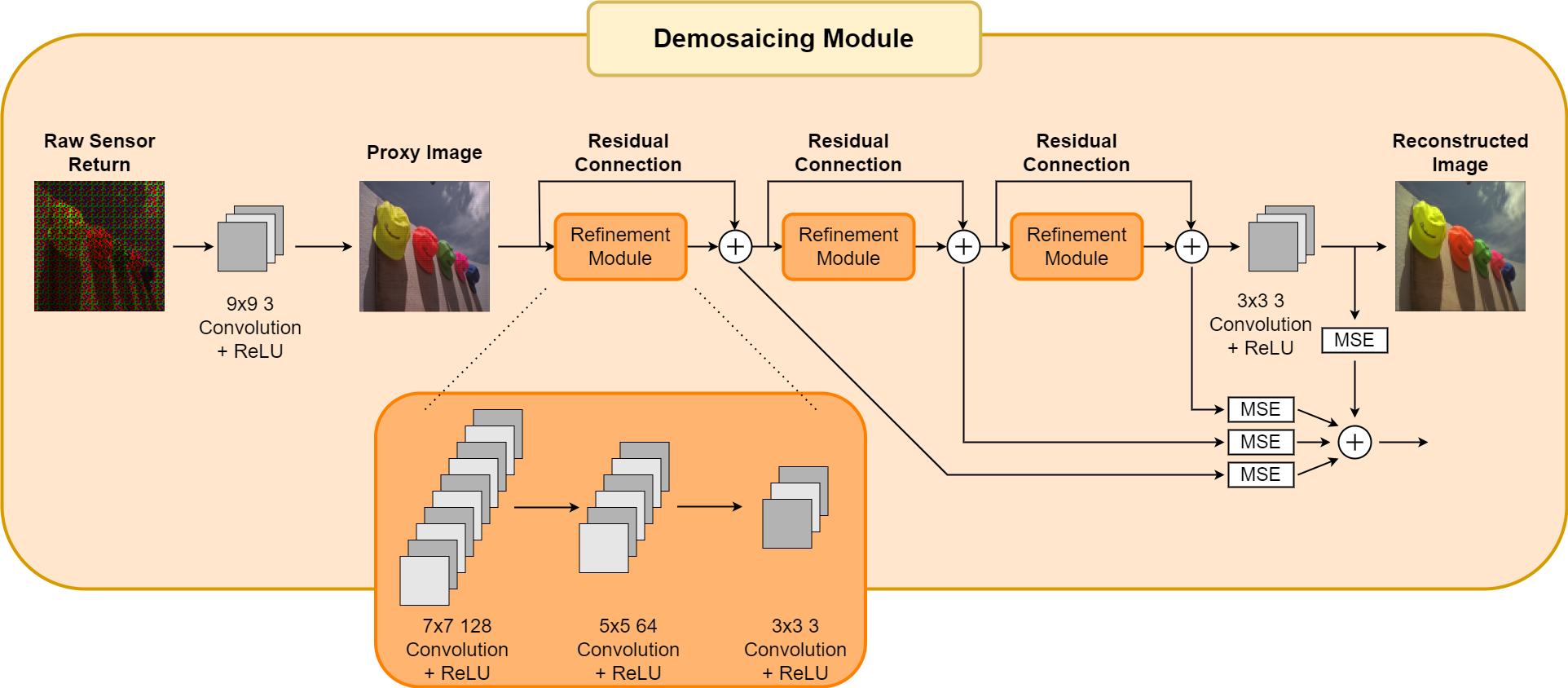}
      \caption{Description of the proposed demosaicer model. The final loss includes intermediate reconstructions.}  
      \label{fig:demosaicer}
      \vspace{-2mm}
\end{figure*}

The HardMax module is designed to be the head of a larger joint network. This makes the HardMax adaptable to any DL-based demosaicing model as a preliminary learnable filter. For this study, we propose a new demosaicing model; a multi-stage convolutional neural network that learns to create a pseudoimage and to refine it through multiple refinement blocks with a cumulative loss function of refinement blocks.

Our demosaicing architecture is illustrated in Figure \ref{fig:demosaicer}. The main structure is designed with ideas from image reconstruction models \cite{shi2019image} and \cite{mdrafi2020joint}. The demosaicing model takes the $3N \times 3N \times C$-size output of HardMax module (representing the sparse raw camera returns) as input and returns a $3N \times 3N \times 3$-size reconstructed RGB image, with the desired $N \times N \times C$ reconstruction in the middle. The reason for the larger input-output patches is to counter the artifact effects that occur at the edges of reconstructed patches, as was also used in \cite{chakrabarti2016learning} and yields good results.

The model first converts the raw camera returns into a pseudo image, similar to \cite{henz2018deep}. But instead of a hand-picked interpolation kernel as in \cite{henz2018deep}, our model creates a pseudo image by applying a convolutional layer with 3 kernels of $9 \times 9$ size with padding for equal-sized output and no bias. The coarse reconstruction is enforced by including the pseudoimage and the consecutive refinement outputs into the loss function as extra terms enforcing them to be similar to the labeled images as well. 

The pseudoimage is then pushed into three consecutive refinement submodules, each of which is composed of three convolutional layers followed by ReLU activation functions. The first layer has 128 kernels of $7 \times 7$ size, the second layer has 64 kernels of $5 \times 5$ size and the third layer has 3 kernels of $3 \times 3$ size. All layers apply padding to preserve the image size and include an $\ell_2$ regularizer. Between each refinement block, there are skip connections added to carry over the gradient. The model's final layer is another convolutional layer with 3 kernels of $3 \times 3$ size and a ReLU activation function. 

The model's final loss function is defined as the sum of all mean squared error (MSE) loss computed over the total $T$ training image patches in a training batch, each refinement module outputs, and the pseudo image using the corresponding label image, as given in \eqref{eq:loss_fun}

\begin{equation}\label{eq:loss_fun}
    \mathcal{L}_{total} = \sum_{i=1}^{T}\sum_{n=0}^{4} (x_i - \hat{x}_i^{(n)})^2,
\end{equation}

\noindent where $x_i$ represents the i-th training sample and $\hat{x}_i^{(n)}$ is the pseudoimage or the n-th refinement module output as shown in Fig.~\ref{fig:demosaicer}. $\hat{x}_i^{(4)}$ denotes the final reconstructed image output of the model. 

In our evaluations, we tested the joint binary CFA learning and demosaicing model with three existing DL-based demosaicing models (explained in Section~\ref{subsec:dl_based_demos}) by only replacing the proposed demosaicing model. In this way, we show that the joint architecture allows the proposed binary CFA learning module to be used with different demosaicing models, and compare the effectiveness of demosaicer models on the learned filter and the overall performance. In addition, any future enhanced DL-based demosaicing model can be integrated with the binary CFA learning module both to learn enhanced CFA filters and achieve better reconstruction results.

\section{Dataset and Training} \label{sec:setup}

\subsection{Dataset and Setup} \label{subsec:dataset}
The training dataset used in this study is created from the training and validation images of the BSDS500 dataset \cite{arbelaez2010contour} totaling 400 images. BSDS500 is a collection of $481 \times 321$ px-sized RGB images originally published for segmentation benchmarks but can be found commonly as training data in other computer vision tasks, including image demosaicing. The training image patches are $3N \times 3N$ non-overlapping side-by-side blocks from this dataset, and the total number of the training patches is $881,600$ in the case of $N=8$ patch size. 

We used two different test image datasets. The first dataset is composed of 20 images selected from BSDS500's test images chosen based on their complexity in contour and texture. The second test dataset used in this study is the Kodak dataset \cite{franzenKodak} which is composed of $24$ different images with $768 \times 512$ sizes. Performance metrics of individual datasets were recorded and compared separately. 

\subsection{Training} \label{subsec:training}
Training and evaluation of all models were conducted with an Nvidia A6000 GPU. The programming language selected for the implementation of the source code is Python 3.10 with TensorFlow library and Keras module. The loss function used in all demosaicing models is the mean squared error (MSE) loss as defined in \eqref{eq:loss_fun}. The Adaptive Movement Estimation (ADAM) optimizer is used during backpropagation \cite{kingma2014adam} with decay rates ($\beta_1$ and $\beta_2$) and division constant ($\hat{\epsilon}$) values of ADAM optimizer were selected as $0.9$, $0.999$ and $10^{-7}$ respectively. The learning rate follows an exponential decay from the value from 0.0001 to 0.00001.

The source code used for training and testing the proposed HardMax model along with the compared models can be found in the \href{https://github.com/msuimpress/hardmax/}{GitHub repository}. 

\subsection{Evaluation} \label{subsec:evaluation}
The performance metrics used in the performance evaluation of the proposed approach are the Peak Signal-to-Noise Ratio (PSNR) and the Structural Similarity Index Metric (SSIM). These metrics are profound in the digital image processing literature for image reconstruction performance evaluations. PSNR is used to observe the distortion of the reconstruction compared to the original signal in terms of decibel units and a higher PSNR value indicates a better reconstruction performance. PSNR can be computed as in \eqref{eq:psnr} where the mean squared error ($MSE$) is computed over the whole reconstructed image.

\begin{equation}\label{eq:psnr}
    PSNR = 10\log_{10} (255^2/{MSE})
\end{equation}

SSIM is used to compare a reconstruction with its ground truth signal in terms of localized similarity. An SSIM score takes a value between 0 and 1, where a higher score indicates that the reconstructed image resembles the original image more. Calculation of SSIM involves the means and variances of the original ($\mu_{x}$, $\sigma_{x}^2$) and the reconstructed ($\mu_{\hat{x}}$, $\sigma_{\hat{x}}^2$) images, along with the covariance of both ($\sigma_{x \hat{x}}$). The division stabilizer constants $c_1$ and $c_2$ in the SSIM definition in \eqref{eq:ssim} are chosen as $(255*0.01)^2$ and $(255*0.03)^2$ respectively.

\begin{equation}\label{eq:ssim}
    SSIM = \frac{(2\mu_{x}\mu_{\hat{x}} + c_1)(2\sigma_{x \hat{x}} + c_2)}{(\mu_{x}^2 +\mu_{\hat{x}}^2+c_1)(\sigma_{x}^2 + \sigma_{\hat{x}}^2 + c_2)}
\end{equation}

\section{Results} \label{sec:results}
The proposed joint binary CFA and demosaicing model is evaluated under six different scenarios. They are briefly summarized as follows: 

\begin{itemize}
    \item The first test in Sec.~\ref{subsec:comp_alt_demos} compares the performance of the proposed demosaicer with alternative demosaicers outlined in Sec.~\ref{subsec:demos}.
    \item  The second test in Sec.~\ref{subsec:comp_filter_size} evaluates the effect of the CFA size in the final reconstruction quality. The learned CFAs for each size under both RGB and RGBW configurations are provided. 
    \item  The third test in Sec.~\ref{subsec:comp_fixed_cfa} compares the performance of the CFA learned by the proposed HardMax module with respect to the popular hand-crafted fixed CFAs such as Bayer, Lukac, and others.
    \item The fourth test in Sec.~\ref{subsec:comp_alt_cfa} compares the performance of the proposed architecture with respect to the alternative CFA learning studies in \cite{chakrabarti2016learning} and \cite{henz2018deep}.
    \item The fifth test in Sec.~\ref{subsec:comp_epoch} analyzes the convergence of the proposed method and the change of the learned CFAs during the training progress. The learned filters at different epochs and the corresponding image reconstructions and performance metrics are provided.
    \item The sixth and final test in Sec.~\ref{subsec:comp_data_size} evaluates the effect of the training data size on both the learned CFAs and the final reconstruction quality under cases from 250 thousand up to 1 million training image patches. 
\end{itemize}

\subsection{Comparison of Demosaicing Models} \label{subsec:comp_alt_demos}
In the first analysis, we compare the performance of the proposed demosaicing model with the existing models \cite{henz2018deep, chakrabarti2016learning, de2021data} presented in Section~\ref{subsec:dl_based_demos}. All demosaicing models are used in the joint framework by keeping the HardMax CFA learning module fixed and changing only the demosaicing models. A total of eight different joint models are trained and tested with the combination of four different demosaicing modules for both color configurations (RGB and RGBW). The average PSNR and SSIM metrics for Kodak and BSDS500 test datasets are provided in Table \ref{tab:table_demos}, with results for RGB and RGBW configurations presented separately.

\begin{table}[!ht]
    \centering
    \begin{subtable}[h]{0.9\textwidth}
        \centering
            \begin{tabular}{ ||c|c||m{3em}|m{3em}|m{3em}|m{3em}||}
            \hline
            \multicolumn{2}{||c||}{Demosaicing Model} & \multicolumn{1}{c|}{in \cite{henz2018deep}} & \multicolumn{1}{c|}{in \cite{chakrabarti2016learning}} & \multicolumn{1}{c|}{in \cite{de2021data}} & \multicolumn{1}{c||}{Proposed} \\
            \hline\hline
            \multirow{2}{4em}{Kodak} & PSNR & 38.321 & 37.738 & 32.252 & \textbf{40.451} \\ \cline{2-6}
            & SSIM & 0.9767 & 0.9714 & 0.9366 & \textbf{0.9844} \\ 
            \hline
            \multirow{2}{4em}{BSDS500} & PSNR & 38.247 & 36.721 & 30.591 & \textbf{40.054} \\ \cline{2-6}
            & SSIM & 0.9853 & 0.9774 & 0.9353 & \textbf{0.9897} \\ 
            \hline
            \end{tabular}
       \caption{RGB}
       \label{tab:table_demos_rgb}
    \end{subtable}
    \\
    \begin{subtable}[h]{0.9\textwidth}
        \centering
            \begin{tabular}{ ||c|c||m{3em}|m{3em}|m{3em}|m{3em}||}
            \hline
            \multicolumn{2}{||c||}{Demosaicing Model} & \multicolumn{1}{c|}{in \cite{henz2018deep}} & \multicolumn{1}{c|}{in \cite{chakrabarti2016learning}} & \multicolumn{1}{c|}{in \cite{de2021data}} & \multicolumn{1}{c||}{Proposed} \\
            \hline\hline
            \multirow{2}{4em}{Kodak} & PSNR & 40.632 & 37.959 & 33.359 & \textbf{41.881} \\ \cline{2-6}
            & SSIM & 0.9863 & 0.9739 & 0.9458 & \textbf{0.9871} \\ 
            \hline
            \multirow{2}{4em}{BSD500} & PSNR & 40.561 & 36.919 & 31.168 & \textbf{41.181} \\ \cline{2-6}
            & SSIM & \textbf{0.9919} & 0.979 & 0.9392 & 0.9918 \\ 
            \hline
            \end{tabular}
        \caption{RGBW}
        \label{tab:table_demos_rgbw}
     \end{subtable}
     \caption{Comparison of different DL-based demosaicing models in \cite{henz2018deep, chakrabarti2016learning, de2021data} used alongside the proposed HardMax CFA module for (a) RGB, and (b) RGBW configurations. The best performance for each metric is shown in bold. }
     \label{tab:table_demos}
\end{table}

The proposed demosaicing architecture shows the highest PSNR and SSIM values for both RGB and RGBW cases. Respectively for Kodak and BSDS500 test datasets, it provides 2.23 dB and 1.81 dB higher PSNR in RGB case, and 1.25 dB and 0.62 dB higher PSNR in RGBW case compared to the second highest demosaicer, the model in \cite{henz2018deep}. For the SSIM metric, the proposed model and the model in \cite{henz2018deep} perform better than the other two, with the proposed model surpassing in most of the test cases.

Categorically we observe that the models with feed-forward fully convolutional architectures (i.e., proposed model and \cite{henz2018deep}) perform comparably better than models with parallel architectures separating the reconstruction of color and texture information such as \cite{chakrabarti2016learning}  and  \cite{de2021data}. The proposed approach adapts deep learning-based image reconstruction to the demosaicing problem and unlike the model in \cite{henz2018deep} that use a hand-crafted kernel for an initial reconstruction, our model rather lets a dedicated convolutional layer learn to reconstruct a pseudo-image, and it uses skip connections and a combined final loss function to reinforce the reconstruction quality. Given the enhanced performance of the proposed demosaicing model, the rest of the evaluation scenarios will use the proposed demosaicing model in the joint architecture along with the HardMax CFA learning module.

\subsection{Analysis on the CFA Size} \label{subsec:comp_filter_size}
In the second analysis, we investigate the effect of the CFA size on reconstruction performance. In this analysis, four different CFA sizes ($4 \times 4$, $8 \times 8$, $12 \times 12$, and $16 \times 16$) have been tested with the proposed joint model under both RGB and RGBW color configurations. The achieved average PSNR and SSIM metrics over the test datasets for varying CFA sizes are presented in Table~\ref{tab:table_cfa_size}, with results for RGB and RGBW configurations presented separately.

\begin{table}[!ht]
    \centering
    \begin{subtable}[h]{0.9\textwidth}
        \centering
            \begin{tabular}{ ||c|c||m{3em}|m{3em}|m{3em}|m{3em}|| } 
            \hline
            \multicolumn{2}{||c||}{CFA Size} & \multicolumn{1}{c|}{4x4} & \multicolumn{1}{c|}{8x8} & \multicolumn{1}{c|}{12x12} & \multicolumn{1}{c||}{16x16} \\
            \hline\hline
            \multirow{2}{4em}{Kodak} & PSNR  & 38.561 & \textbf{40.451} & 39.535 & 38.865 \\ \cline{2-6}
            & SSIM & 0.9777 & \textbf{0.9844} & 0.9824 & 0.9813 \\ 
            \hline
            \multirow{2}{4em}{BSD500} & PSNR  & 37.57 & \textbf{40.054} & 38.676 & 37.880 \\ \cline{2-6}
            & SSIM & 0.9826 & \textbf{0.9897} & 0.9873 & 0.9856 \\ 
            \hline
            \end{tabular}
       \caption{RGB}
       \label{tab:table_cfa_size_rgb}
    \end{subtable}
    \\
    \begin{subtable}[h]{0.9\textwidth}
        \centering
            \begin{tabular}{ ||c|c||m{3em}|m{3em}|m{3em}|m{3em}|| } 
            \hline
            \multicolumn{2}{||c||}{CFA Size} & \multicolumn{1}{c|}{4x4}  & \multicolumn{1}{c|}{8x8} & \multicolumn{1}{c|}{12x12} & \multicolumn{1}{c||}{16x16} \\
            \hline\hline
            \multirow{2}{4em}{Kodak} & PSNR  & 40.189 & \textbf{41.881} & 39.89 & 38.988 \\ \cline{2-6}
            & SSIM  & 0.9842  & \textbf{0.9871} & 0.9825 & 0.9814 \\ 
            \hline
            \multirow{2}{4em}{BSD500} & PSNR & 39.700 & \textbf{41.181} & 38.59 & 37.878 \\ \cline{2-6}
            & SSIM & 0.9891 & \textbf{0.9918} & 0.987 & 0.9848 \\ 
            \hline
            \end{tabular}
        \caption{RGBW}
        \label{tab:table_cfa_size_rgbw}
     \end{subtable}
     \caption{Comparison of different CFA sizes for (a) RGB and (b) RGBW configurations.}
     \label{tab:table_cfa_size}
     \vspace{-2mm}
\end{table}

Our first observation from the Table~\ref{tab:table_cfa_size} is that the highest PSNR and SSIM performance is achieved for $8 \times 8$ CFA size in both RGB and RGBW color configurations. Using smaller or larger CFA sizes than $8 \times 8$ affects the final reconstruction negatively. Another important point is that the existence of a luminance channel boosts the final performance, leading to higher PSNR and SSIM results. This means that color interpolation with sparse representation is possible if the luminance information is present.

\begin{figure}[!ht]
    \centering
    \begin{subfigure}{0.23\textwidth}
        \centering
        \includegraphics[width=0.95\textwidth]{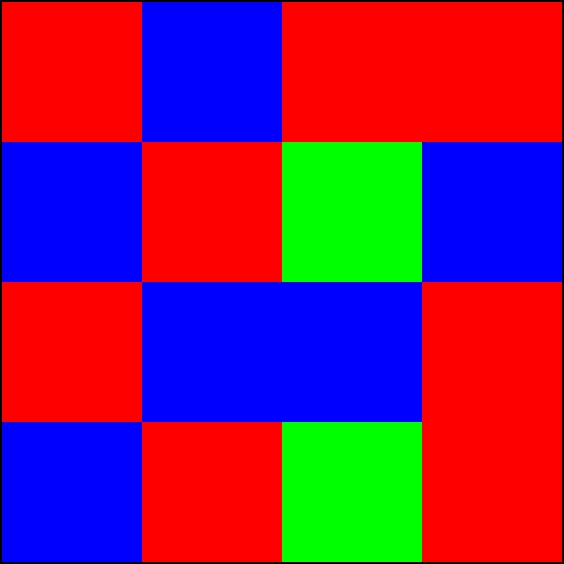}
        \caption{}
    \end{subfigure}
    \begin{subfigure}{0.23\textwidth}
        \centering
        \includegraphics[width=0.95\textwidth]{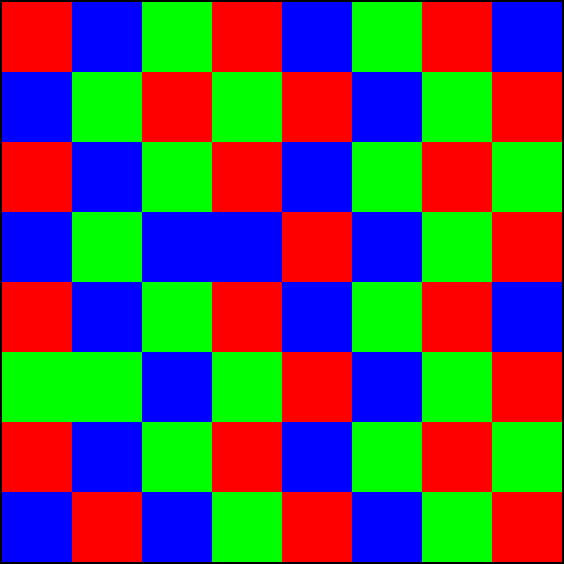}
        \caption{}
    \end{subfigure}
    \begin{subfigure}{0.23\textwidth} 
        \centering
        \includegraphics[width=0.95\textwidth]{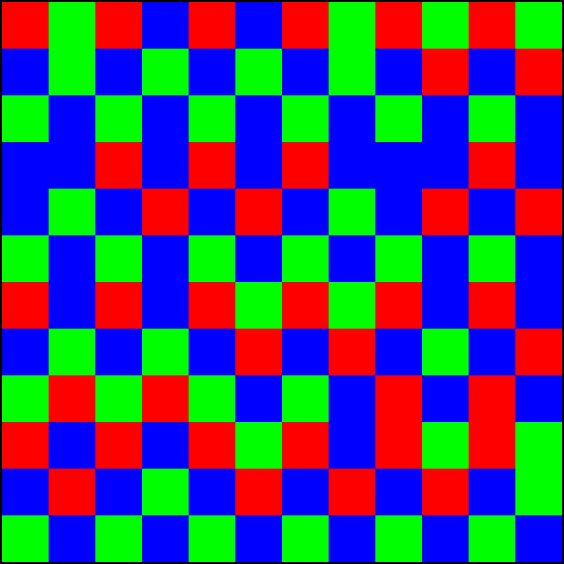}
        \caption{}
    \end{subfigure}
    \begin{subfigure}{0.23\textwidth}
        \centering
        \includegraphics[width=0.95\textwidth]{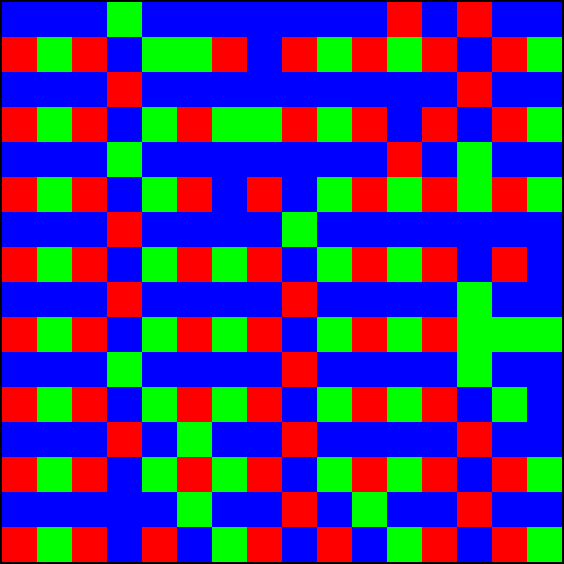}
        \caption{}
    \end{subfigure}
    \\
    \begin{subfigure}{0.23\textwidth}
        \centering
        \includegraphics[width=0.95\textwidth]{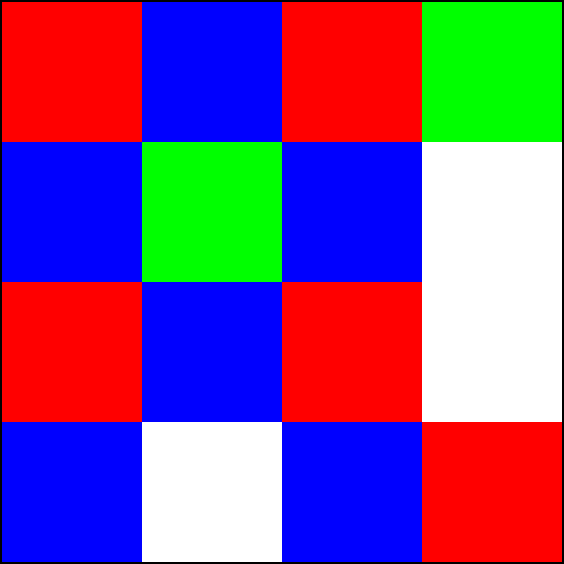}
        \caption{}
    \end{subfigure}
    \begin{subfigure}{0.23\textwidth}
        \centering
        \includegraphics[width=0.95\textwidth]{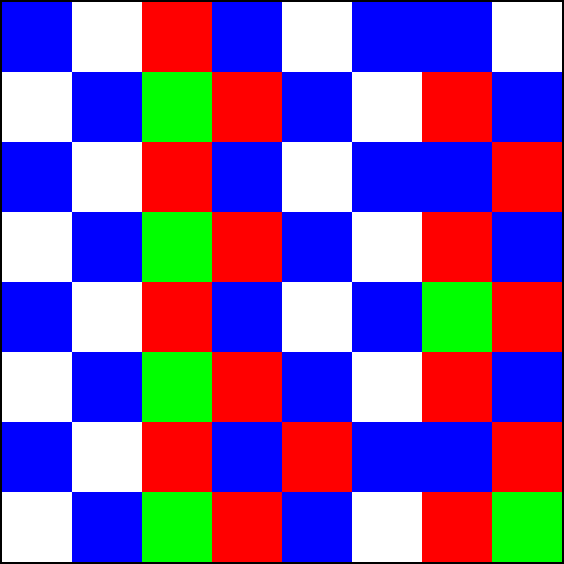}
        \caption{}
    \end{subfigure}
    \begin{subfigure}{0.23\textwidth}
        \centering
        \includegraphics[width=0.95\textwidth]{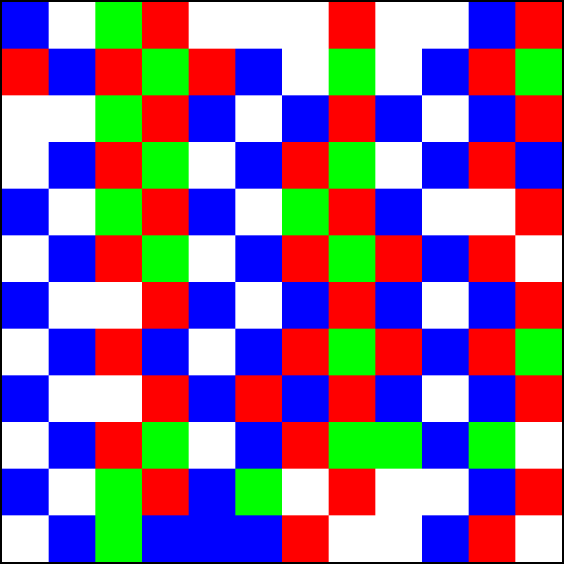}
        \caption{}
    \end{subfigure}
    \begin{subfigure}{0.23\textwidth}
        \centering
        \includegraphics[width=0.95\textwidth]{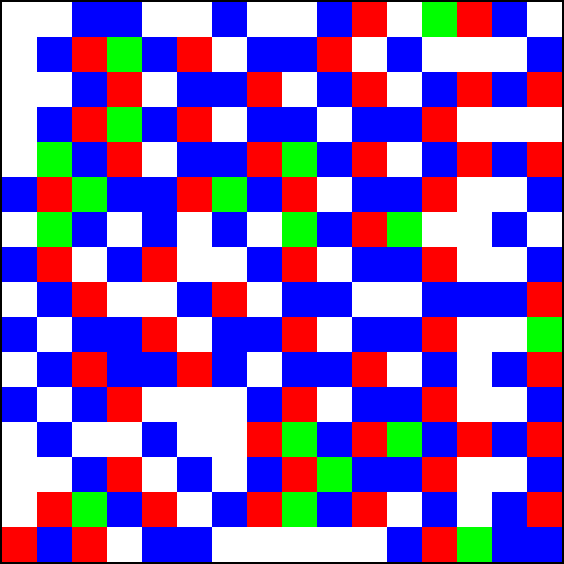}
        \caption{}
    \end{subfigure}
    \caption{CFAs with different sizes learned by the joint HardMax CFA optimization and the proposed demosaicing model. (a) $4 \times 4$, (b) $8 \times 8$, (c) $12 \times 12$, (d) $16 \times 16$ filters for RGB configuration; (e) $4 \times 4$, (f) $8 \times 8$, (g) $12 \times 12$, (h) $16 \times 16$ filters for RGBW configuration.}
    \label{fig:scenario_2_learned_cfas}
    \vspace{-2mm}
\end{figure}

The learned CFAs for each different size are visualized in Fig.~\ref{fig:scenario_2_learned_cfas} for both RGB and RGBW configurations. While learned filters don't exactly match with any existing fixed CFA pattern, they seem to show some level of regularity in sampling a color channel. One observation is that the learned CFAs seem to prioritize the blue color over the others in most instances. The second most common channel is red for RGB and luminance for RGBW cases. This is an interesting contrast to the intuition of the hand-crafted CFAs such as Bayer, Lukac, and X-Trans where the green channel is prioritized with the assumption that the human eye is more sensitive to green channel information. While the inherent features of digital images might have enforced the CFA learning module to pick up blue and white channels to minimize the defined mean squared loss term, we believe it is important to understand the reasons DL models are making these selections.

It is also important to note that this study deals particularly with the image reconstruction problem, hence the learned CFAs are objective-specific and trained and optimized for the reconstruction task. Therefore the HardMax CFA learning module combined with a DNN model for a different objective may result in prioritizing to select different channels. 

\subsection{Comparison of Learned CFAs with Hand-Crafted CFAs} \label{subsec:comp_fixed_cfa}
In the third analysis we compare the performance of the CFAs learned by the proposed HardMax approach with the hand-crafted fixed CFAs using the same proposed demosaicer model. For this test, two RGB (Bayer and Lukac) and two RGBW (RGBW and CFZ) filters, shown in Figure~\ref{fig:cfa_examples}, were selected. For each fixed CFA, training image patches are filtered with the respective filter beforehand and the proposed demosaicing model is trained over the filtered training dataset separately for each CFA type. All cases are evaluated with $8 \times 8$ CFA size. The achieved average PSNR and SSIM metrics over the two test datasets can be found in Table~\ref{tab:table_fixed_cfa_comparison}.

\begin{table}[!ht]
    \begin{subtable}[h]{0.48\textwidth}
        \centering
        \begin{tabular}{ ||c|c||m{3em}|m{3em}|m{3.5em}|| } 
            \hline
            \multicolumn{2}{||c||}{CFA} & Bayer & Lukac & Proposed \\
            \hline\hline
            \multirow{2}{4em}{Kodak} & PSNR & 39.036 & 38.683 & \textbf{40.451} \\ \cline{2-5}
            & SSIM & 0.9786 & 0.9808 & \textbf{0.9844} \\ 
            \hline
            \multirow{2}{4em}{BSD500} & PSNR & 39.584 & 38.897 & \textbf{40.054} \\ \cline{2-5}
            & SSIM & 0.9877 & 0.9873 & \textbf{0.9897} \\ 
            \hline
        \end{tabular}
        \caption{RGB}
        \label{tab:table_fixed_cfa_comparison_rgb}
    \end{subtable}
    \hfill
    \begin{subtable}[h]{0.48\textwidth}
        \centering
        \begin{tabular}{ ||c|c||m{3em}|m{3em}|m{3.5em}|| } 
            \hline
            \multicolumn{2}{||c||}{CFA} & RGBW & CFZ & Proposed \\
            \hline\hline
             \multirow{2}{4em}{Kodak} & PSNR & 41.051 & 38.643 & \textbf{41.881} \\ \cline{2-5}
            & SSIM & 0.9866 & 0.9809 & \textbf{0.9871} \\ 
            \hline
            \multirow{2}{4em}{BSD500} & PSNR & 40.820 & 37.981 & \textbf{41.181} \\ \cline{2-5}
            & SSIM & 0.9913 & 0.9859 & \textbf{0.9918} \\ 
            \hline
        \end{tabular}
        \caption{RGBW}
        \label{tab:table_fixed_cfa_comparison_rgbw}
    \end{subtable}
     \caption{Comparison between the performance of the HardMax CFAs and hand-crafted CFAs with the proposed demosaicing model. For (a) RGB and (b) RGBW configurations.}
     \label{tab:table_fixed_cfa_comparison}
\end{table}

The results show that the proposed demosaicer with the learned HardMax CFA shows the highest reconstruction performance in both RGB and RGBW configurations for both test datasets. Considering that for each CFA the demosaicing model is same, this analysis shows that the demosaicer trained together with a learned CFA surpasses the hand-crafted CFAs in reconstruction quality. This is important in showing that filter learning incorporated into the demosaicing in a single training pipeline exploits the features of the training data for the best reconstruction compared to a general work-for-all fixed CFA.

\begin{table}[!ht]
    \centering
    \begin{subtable}[h]{0.9\textwidth}
        \centering
        \begin{tabular}{ ||c|c||m{5.5em}||m{3.5em}|m{3.5em}||} 
            \hline
            \multicolumn{2}{||c||}{CFA Learning Module} & Unconstrained & \multicolumn{2}{|c||}{Constrained}\\ \cline{3-5}
            \multicolumn{2}{||c||}{} &  Linear & Weighted SoftMax & HardMax \\
            \hline\hline
            \multirow{2}{4em}{Kodak} & PSNR & 49.435 & 39.034 & 40.451 \\ \cline{2-5}
            & SSIM & 0.9987 & 0.9782 & 0.9844 \\ 
            \hline
            \multirow{2}{4em}{BSD500} & PSNR & 49.426 & 37.819 & 40.054 \\ \cline{2-5}
            & SSIM & 0.9991 & 0.9637 & 0.9897 \\ 
            \hline
        \end{tabular}
        \caption{}
        \label{tab:table_alt_cfa_comparison_rgb}
    \end{subtable}
    \\
    \begin{subtable}[h]{0.9\textwidth}
        \centering
        \begin{tabular}{ ||c|c||m{5.5em}||m{3.5em}|m{3.5em}||} 
            \hline
            \multicolumn{2}{||c||}{CFA Learning Module} & Unconstrained & \multicolumn{2}{|c||}{Constrained}\\ \cline{3-5}
            \multicolumn{2}{||c||}{} &  Linear & Weighted SoftMax & HardMax \\
            \hline\hline
            \multirow{2}{4em}{Kodak} & PSNR & 49.389 & 39.407 & 41.881 \\ \cline{2-5}
            & SSIM & 0.9982 & 0.9788 & 0.9871 \\ 
            \hline
            \multirow{2}{4em}{BSD500} & PSNR & 49.907 & 38.074 & 41.181 \\ \cline{2-5}
            & SSIM & 0.9989 & 0.9839 & 0.9918 \\ 
            \hline
        \end{tabular}
        \caption{}
        \label{tab:table_alt_cfa_comparison_rgbw}
    \end{subtable}
     \caption{The performance of tested CFA learning models (a) for RGB  and (b)RGBW configurations.}
     \label{tab:table_alt_cfa_comparison}
\end{table}

\subsection{Comparison with Alternative CFA Learning Methods} \label{subsec:comp_alt_cfa}
In this analysis, we compare the performance of the proposed HardMax CFA learning module with the state-of-the-art DL-based CFA learning approaches \cite{henz2018deep, chakrabarti2016learning} which are summarized in Section \ref{subsec:cfa_learning}. The approach in \cite{henz2018deep} learns unconstrained linear weights for each pixel rather than constraining the solution space to only binary weights or single channel selection at each pixel as in the proposed HardMax method and the Weighted SoftMax in\cite{chakrabarti2016learning}. It is expected that learning unconstrained linear weights will provide better performance since it has a much wider optimization space. However, we compared all three approaches over the same training and test dataset combinations. All the alternative joint CFA learning-demosaicing methods were trained and tested as they were proposed in their respective studies. The achieved PSNR and SSIM results can be found in Table~\ref{tab:table_alt_cfa_comparison} for both RGB and RGBW configurations. 

\begin{figure*}[!ht]
    \centering
    \begin{subfigure}{0.24\textwidth}
        \centering
        \includegraphics[width=0.95\textwidth]{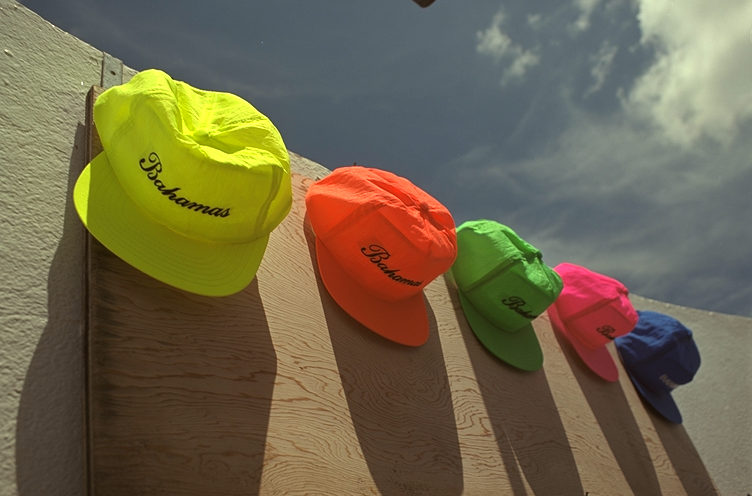}
        \caption{Original}
    \end{subfigure}
    \begin{subfigure}{0.24\textwidth} 
        \centering
        \includegraphics[width=0.95\textwidth]{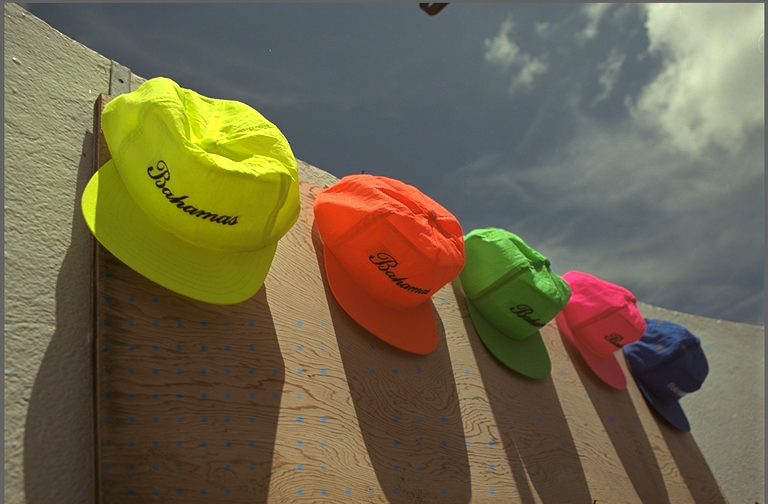}
        \caption{PSNR: 39.0359}
    \end{subfigure}
    \begin{subfigure}{0.24\textwidth}
        \centering
        \includegraphics[width=0.95\textwidth]{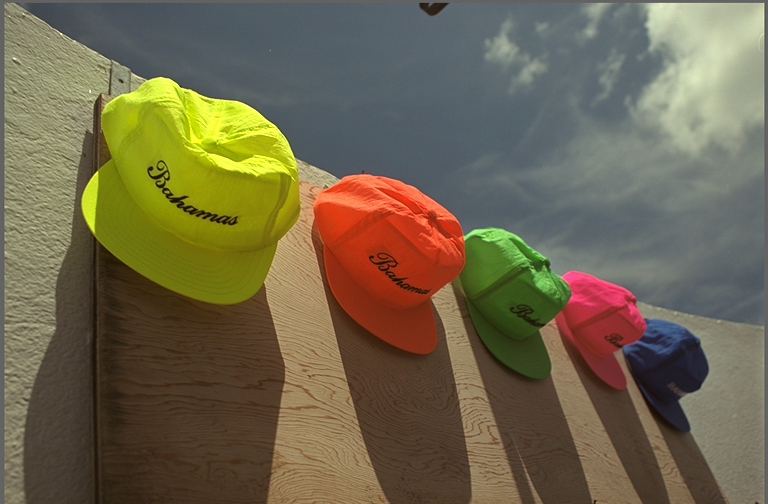}
        \caption{PSNR: 39.6394}
    \end{subfigure}
    \begin{subfigure}{0.24\textwidth}
        \centering
        \includegraphics[width=0.95\textwidth]{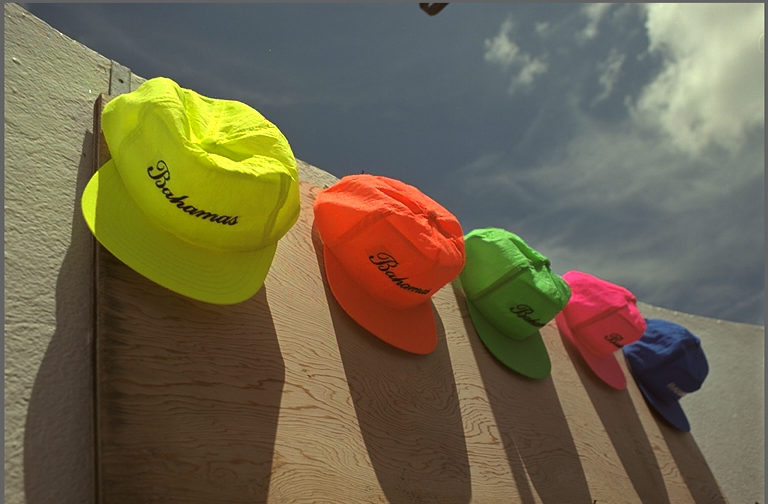}
        \caption{PSNR: 42.5311}
    \end{subfigure}
    \\
    \begin{subfigure}{0.24\textwidth}
        \centering
        \includegraphics[width=0.95\textwidth]{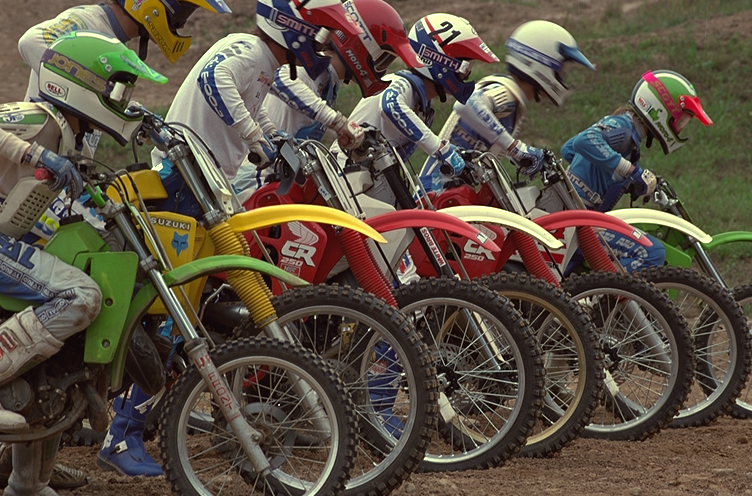}
        \caption{Original}
    \end{subfigure}
    \begin{subfigure}{0.24\textwidth} 
        \centering
        \includegraphics[width=0.95\textwidth]{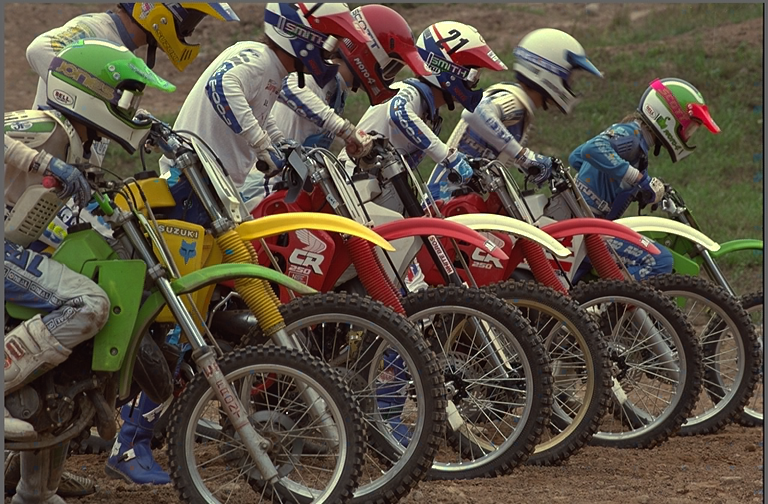}
        \caption{PSNR: 37.7100}
    \end{subfigure}
    \begin{subfigure}{0.24\textwidth}
        \centering
        \includegraphics[width=0.95\textwidth]{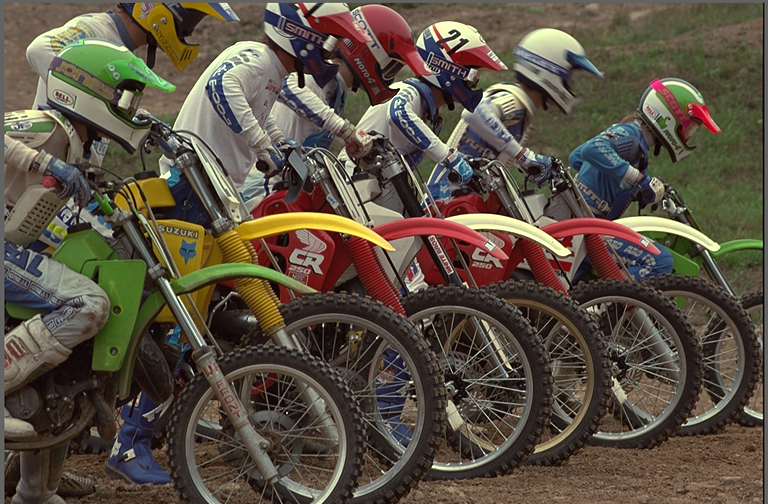}
        \caption{PSNR: 38.7221}
    \end{subfigure}
    \begin{subfigure}{0.24\textwidth}
        \centering
        \includegraphics[width=0.95\textwidth]{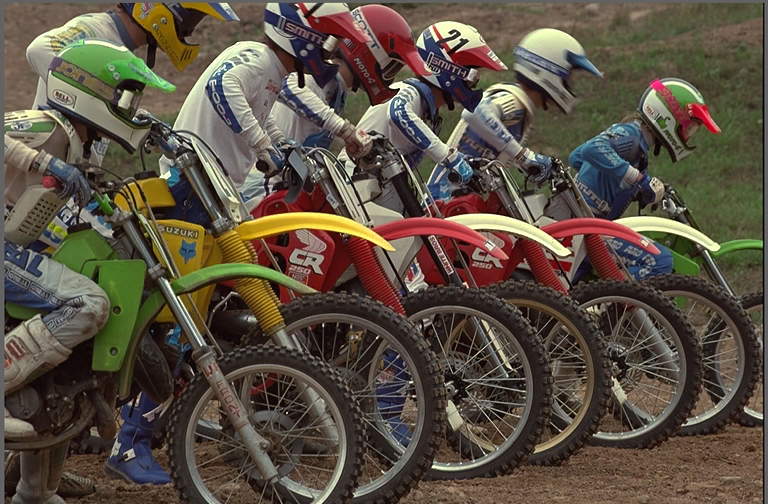}
        \caption{PSNR: 39.4920}
    \end{subfigure}
    \\
    \begin{subfigure}{0.24\textwidth}
        \centering
        \includegraphics[width=0.95\textwidth]{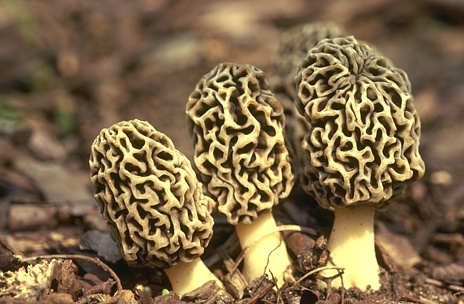}
        \caption{Original}
    \end{subfigure}
    \begin{subfigure}{0.24\textwidth} 
        \centering
        \includegraphics[width=0.95\textwidth]{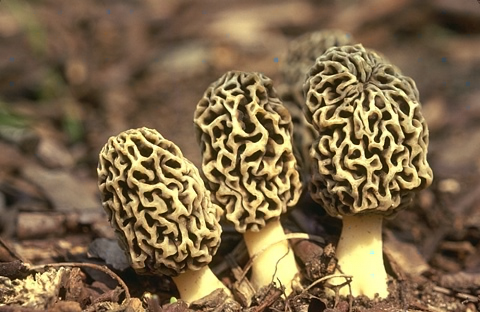}
        \caption{PSNR: 38.9645}
    \end{subfigure}
    \begin{subfigure}{0.24\textwidth}
        \centering
        \includegraphics[width=0.95\textwidth]{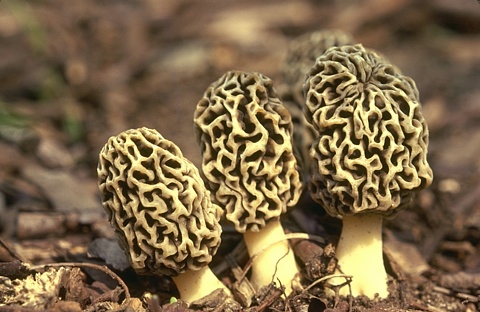}
        \caption{PSNR: 39.7484}
    \end{subfigure}
    \begin{subfigure}{0.24\textwidth}
        \centering
        \includegraphics[width=0.95\textwidth]{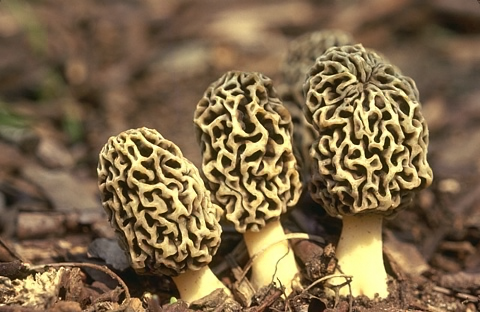}
        \caption{PSNR: 40.3756}
    \end{subfigure}
    \\
    \begin{subfigure}{0.24\textwidth}
        \centering
        \includegraphics[width=0.95\textwidth]{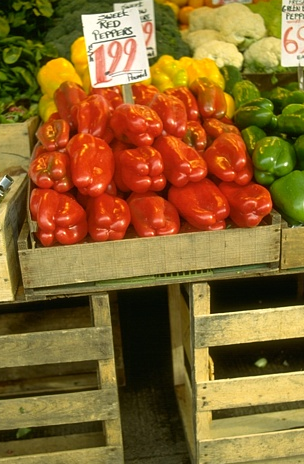}
        \caption{Original}
    \end{subfigure}
    \begin{subfigure}{0.24\textwidth}
        \centering
        \includegraphics[width=0.95\textwidth]{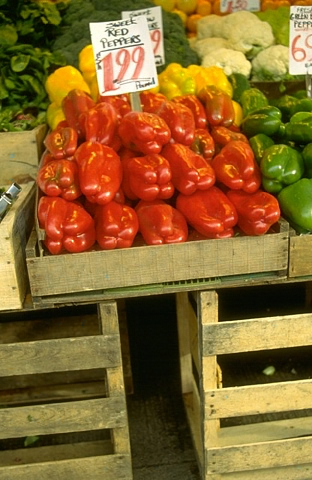}
        \caption{PSNR: 40.2638}
    \end{subfigure}
    \begin{subfigure}{0.24\textwidth}
        \centering
        \includegraphics[width=0.95\textwidth]{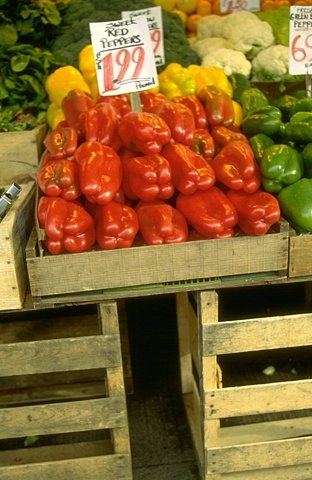}
        \caption{PSNR: 39.3830}
    \end{subfigure}
    \begin{subfigure}{0.24\textwidth}
        \centering
        \includegraphics[width=0.95\textwidth]{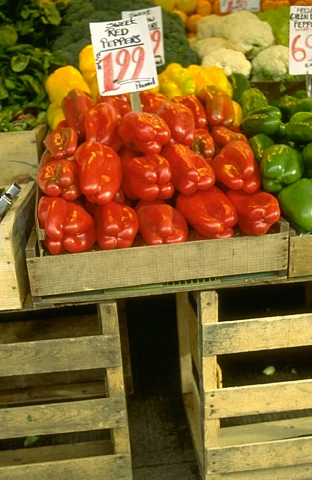}
        \caption{PSNR: 40.4574}
    \end{subfigure}
    \caption{Example reconstructed images in RGB case. From left to right: Original images, and reconstructed images with Bayer, Weighted SoftMax, and the proposed HardMax CFAs. The PSNR value for each reconstructed image is shown below the image.}
    \label{fig:scenario_4_pics}
\end{figure*}

\begin{figure*}[!ht]
    \centering
    \begin{subfigure}{0.24\textwidth}
        \centering
        \includegraphics[width=0.95\textwidth]{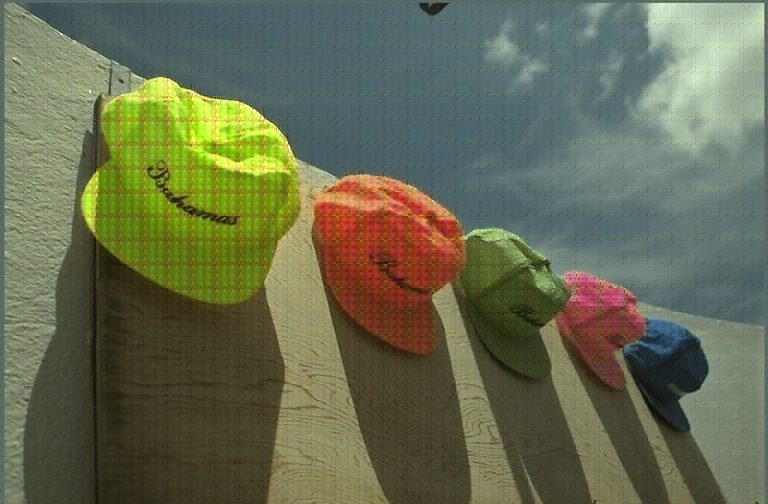}
        \caption{Epoch 1, PSNR: 26.9470}
    \end{subfigure}
    \begin{subfigure}{0.24\textwidth} 
        \centering
        \includegraphics[width=0.95\textwidth]{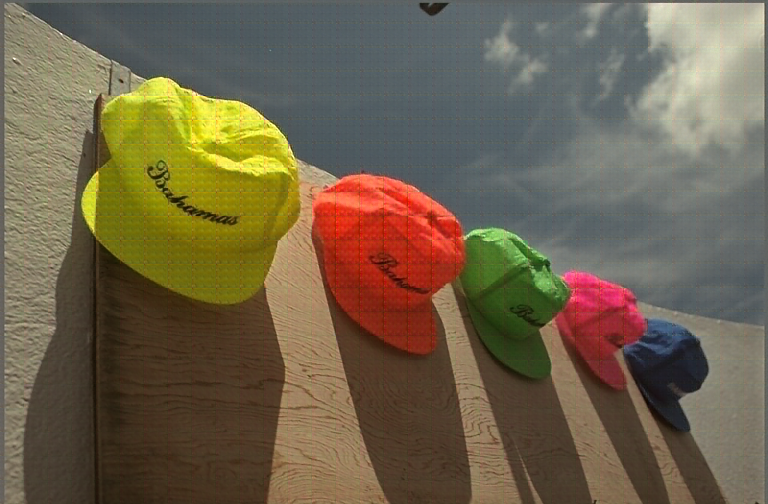}
        \caption{Epoch 2, PSNR: 32.0163}
    \end{subfigure}
    \begin{subfigure}{0.24\textwidth}
        \centering
        \includegraphics[width=0.95\textwidth]{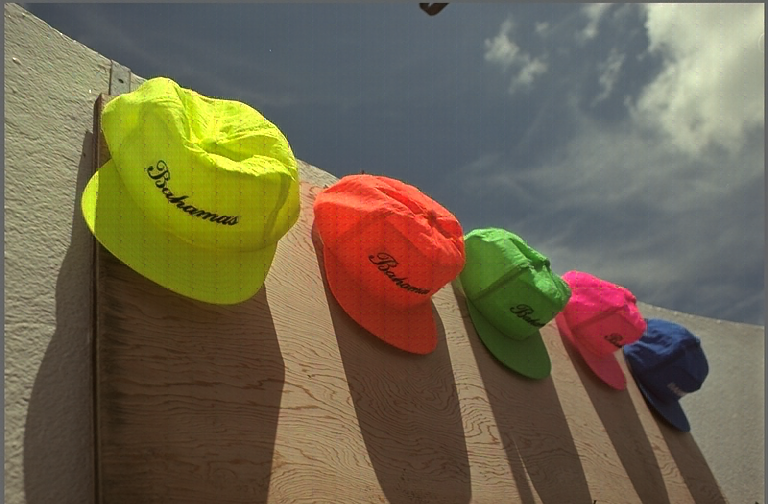}
        \caption{Epoch 4, PSNR: 36.7329}
    \end{subfigure}
    \begin{subfigure}{0.24\textwidth}
        \centering
        \includegraphics[width=0.95\textwidth]{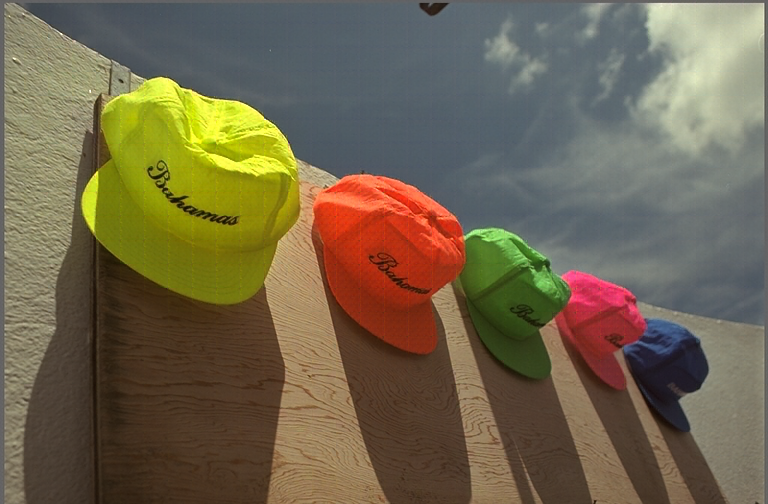}
        \caption{Epoch 5, PSNR: 38.8575}
    \end{subfigure}
    \\
    \begin{subfigure}{0.24\textwidth}
        \centering
        \includegraphics[width=0.95\textwidth]{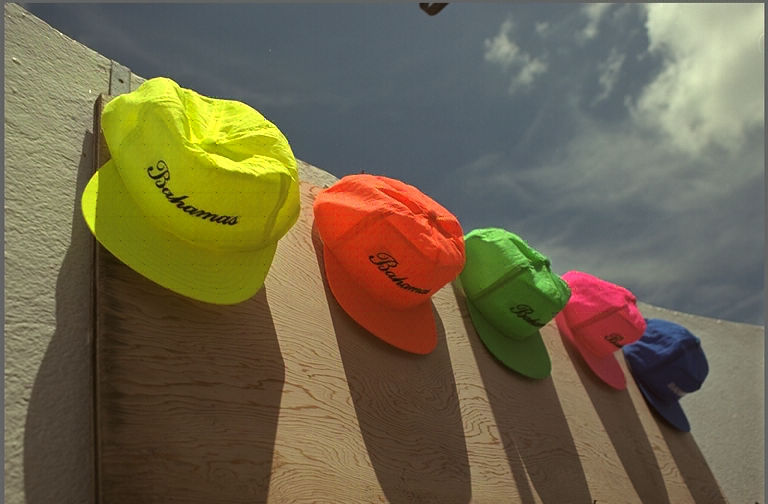}
        \caption{Epoch 10, PSNR: 40.8191}
    \end{subfigure}
    \begin{subfigure}{0.24\textwidth} 
        \centering
        \includegraphics[width=0.95\textwidth]{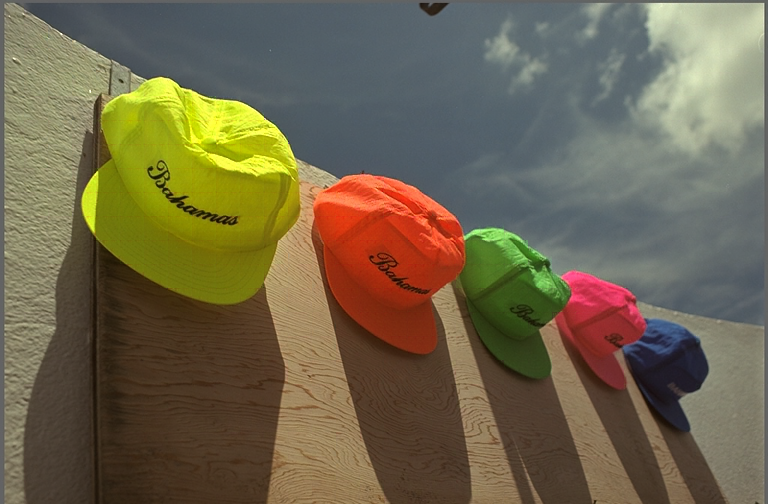}
        \caption{Epoch 15, PSNR: 41.8566}
    \end{subfigure}
    \begin{subfigure}{0.24\textwidth}
        \centering
        \includegraphics[width=0.95\textwidth]{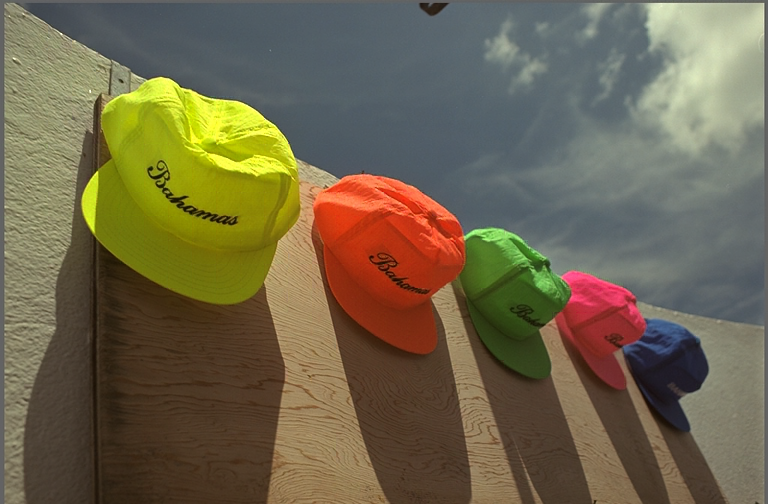}
        \caption{Epoch 20, PSNR: 42.0437}
    \end{subfigure}
    \begin{subfigure}{0.24\textwidth}
        \centering
        \includegraphics[width=0.95\textwidth]{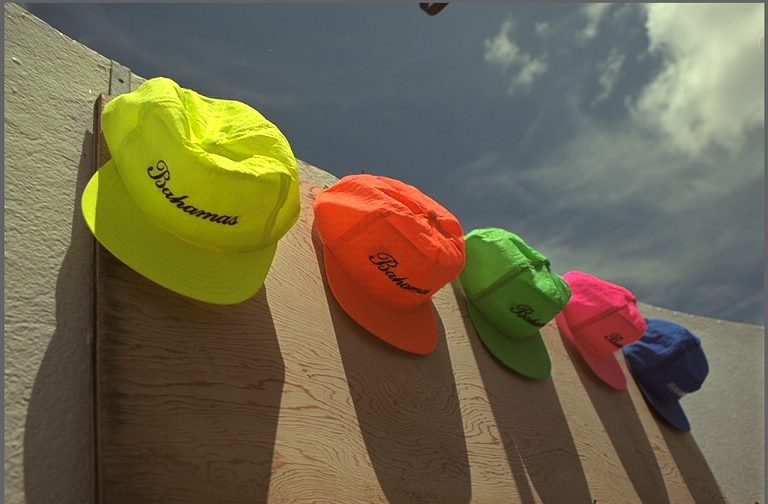}
        \caption{Epoch 50, PSNR: 42.4272}
    \end{subfigure}
    \caption{A sample reconstructed image with RGBW CFA with PSNR values as the training progresses, at (a) Epoch 1, (b) Epoch 2, (c) Epoch 4, (d) Epoch 5, (e) Epoch 10, (f) Epoch 15, (g) Epoch 20, (h) Epoch 50.}
    \label{fig:scenario_5_pics}
    \vspace{1mm}
\end{figure*}

The unconstrained linear CFA outperforms the constrained binary CFAs in all compared metrics under both RGB and RGBW configurations. Linear filter in \cite{henz2018deep} was expected to outperform both binary filters since it has a larger optimization space and is not bound to selecting one single color channel per pixel. If it is possible to construct a camera that can acquire measurements at each pixel as linear weighted combination of each color channel with weights learned through the CFA learning it would be optimal. However, for most practical cameras, which are constrained to observe only one of the color channels for each pixel at a time, we can compare the performance of the proposed approach and the Weighted SoftMax in \cite{chakrabarti2016learning}. This comparison can be found under the constrained column of Table~\ref{tab:table_alt_cfa_comparison} and it can be seen that the proposed HardMax CFA learning approach surpasses the Weighted SoftMax by 1.42 dB in RGB case and 1.48 dB in RGBW case and also results in higher percentage values for the SSIM scores for both test datasets. 

We illustrate four example test images chosen from the two test datasets in Fig.~\ref{fig:scenario_4_pics} along with their reconstructions from fixed Bayer CFA, and learned CFAs from Weighted SoftMax and the proposed HardMax approach. The achieved PSNR values for each individual image are shown in the same figure and it can be seen that the proposed joint architecture results in higher PSNR values than fixed Bayer or Weighted SoftMax based learned CFAs. In addition to the PSNR metric, visual comparisons can be done on Fig.~\ref{fig:scenario_4_pics} between the original images and the reconstructions from different CFAs.  

\subsection{Analysis on the Learned CFA in Training}\label{subsec:comp_epoch}
As we establish the high performance of the proposed joint binary CFA learning and the demosaicing model, in this analysis we start looking into the behavior of the model during the training process. For this analysis, we observe the effect of the training by comparing the performance metrics, reconstruction qualities, and learned filters at different epochs as the training progresses. In this test, one sample of the proposed joint model was trained, and its weights were saved after epochs 1, 2, 4, 5, 10, 15, 20, and 50. These saved weights were used with the test dataset to evaluate the change in image reconstruction quality. 

\begin{table}[!ht]
    \centering
    \begin{subtable}[h]{1\textwidth}
        \centering
        \begin{tabular}{||c|c||m{3.5em}|m{3.5em}|m{3.5em}|m{3.5em}|m{3.5em}|m{3.5em}|m{3.5em}|m{3.5em}||} 
            \hline
            \multicolumn{2}{||c||}{Epoch} & 1 & 2 & 4 & 5 & 10 & 15 & 20 & 50 \\
            \hline\hline
            \multirow{2}{4em}{Kodak} & PSNR & 27.2751 & 31.6592 & 35.9546 & 37.4252 & 39.2274 & 40.0207 & 39.9918 & 40.4209 \\ \cline{2-10}
            & SSIM & 0.8406 & 0.8967 & 0.9591 & 0.9677 & 0.9795 & 0.983 & 0.9833 & 0.984 \\ 
            \hline
            \multirow{2}{4em}{BSDS500} & PSNR & 26.6396 & 30.2478 & 34.5898 & 36.3041 & 38.2320 & 39.1354 & 39.3217 & 39.8671 \\ \cline{2-10}
            & SSIM & 0.8557 & 0.907 & 0.9624 & 0.9737 & 0.9836 & 0.9869 & 0.9878 & 0.9887 \\ 
            \hline
        \end{tabular}
        \caption{}
        \label{tab:table_alt_cfa_epoch_comp_rgb}
    \end{subtable}
    \\
    \begin{subtable}[h]{1\textwidth}
        \centering
        \begin{tabular}{||c|c||m{3.5em}|m{3.5em}|m{3.5em}|m{3.5em}|m{3.5em}|m{3.5em}|m{3.5em}|m{3.5em}||} 
            \hline
            \multicolumn{2}{||c||}{Epoch} & 1 & 2 & 4 & 5 & 10 & 15 & 20 & 50 \\
            \hline\hline
            \multirow{2}{4em}{Kodak} & PSNR & 27.9818 & 26.8873 & 37.4772 & 38.5111 & 39.7669 & 39.7566 & 40.6076 & 40.6922 \\ \cline{2-10}
            & SSIM & 0.8312 & 0.874 & 0.9725 & 0.9747 & 0.9819 & 0.9843 & 0.9853 & 0.9858 \\ 
            \hline
            \multirow{2}{4em}{BSDS500} & PSNR & 27.9691 & 27.0902 & 37.3992 & 38.4566 & 40.0099 & 40.2789 & 41.0271 & 41.3370 \\ \cline{2-10}
            & SSIM & 0.8497 & 0.879 & 0.9775 & 0.9803 & 0.9866 & 0.9889 & 0.9899 & 0.9908 \\ 
            \hline
        \end{tabular}
        \caption{}
        \label{tab:table_alt_cfa_epoch_comp_rgbw}
    \end{subtable}
    \caption{Performance of the proposed model during training after different number of epochs. For (a) RGB and (b) RGBW configurations.}
    \label{tab:table_epoch_comp}
\end{table}

Table~\ref{tab:table_epoch_comp} shows the performance of the model at different epochs. The results show that both PSNR and SSIM metrics improve with increasing epochs and converge to their final values. The PSNR after a single epoch is at 27dB, and after epoch 20 this result was improved to approximately 40dB and the final average PSNR at epoch 50 is 40.4dB over the test dataset. As the training progresses, the model does a better job at reconstructing the images while the CFA pattern is also forming in training. Figure \ref{fig:scenario_5_pics} shows an example reconstructed image at different epoch numbers. While the reconstruction artifacts are clearly visible at early epochs, after 15 to 20 epochs, image reconstruction performance improves, and finally a $42.42$ dB PSNR is achieved after epoch 50. 

\subsection{Effect of Training Dataset Size on Learned CFAs} \label{subsec:comp_data_size}
In the sixth and final analysis, our goal is to understand the effect of the training data size on the learned CFA and the performance of the demosaicing model. We created training datasets with $250$K, $500$K, and $1$ Million samples of $8 \times 8$ size image blocks. For each dataset size, the proposed model is trained 10 independent times with random weight initialization. 

Table~\ref{tab:table_avg_filters} shows the mean and variance in the PSNR and SSIM metrics over Kodak and BSDS500 datasets. As the size of the training dataset increases, both the average PSNR and SSIM metrics improve. Another important observation is that the variance in the achieved performances decreases as models are trained over larger datasets. This shows that even though each training might end up with a different learned CFA and a demosaicing model, the achieved performances with these models are mostly consistent with lower variance in larger datasets. 

\begin{table}[ht!]
    \centering
    \begin{tabular}{||c|c||m{3em}|m{3em}|m{3em}|m{3em}|m{3em}|m{3em}||} 
        \hline
        \multicolumn{2}{||c||}{\multirow{2}{4em}{Dataset Size}} & \multicolumn{2}{c|}{1 Mil.} & \multicolumn{2}{|c|}{500K} & \multicolumn{2}{|c||}{250K} \\ \cline{3-8} 
        \multicolumn{2}{||c||}{} & Avg & Var & Avg & Var & Avg & Var \\ 
        \hline\hline
        \multirow{2}{4em}{Kodak} & PSNR & \textbf{40.357} & 0.0297 & 39.866 & 0.1423 & 39.100 & 0.2218 \\ 
        \cline{2-8} & SSIM & \textbf{0.985} & 0.00196 & 0.983 & 0.00139 & 0.97 & 0.00357 \\ 
        \hline
        \multirow{2}{4em}{BSDS500} & PSNR & \textbf{39.843} & 0.1330 & 39.217 & 0.2332 & 38.245 & 0.2967 \\ 
        \cline{2-8} & SSIM & \textbf{0.989} & 0.00138 & 0.988 & 0.00182 & 0.984 & 0.00333 \\ 
        \hline
    \end{tabular}
    \vspace{2mm}
    \caption{The mean and variance in the PSNR and SSIM metrics over Kodak and BSDS500 test images for various training dataset sizes over 10 independent trainings with random weight initialization.}
    \label{tab:table_avg_filters}
\end{table}

\begin{figure}[ht!]
    \centering
    \begin{subfigure}{0.24\textwidth}
        \centering
        \includegraphics[width=0.95\textwidth]{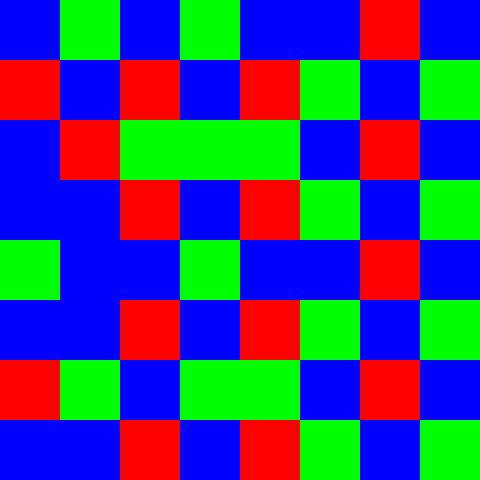}
        \caption{}
    \end{subfigure}
    \begin{subfigure}{0.24\textwidth}
        \centering
        \includegraphics[width=0.95\textwidth]{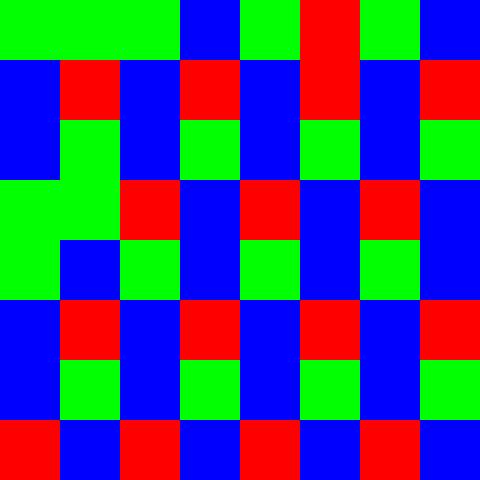}
        \caption{}
    \end{subfigure}
    \begin{subfigure}{0.24\textwidth}
        \centering
        \includegraphics[width=0.95\textwidth]{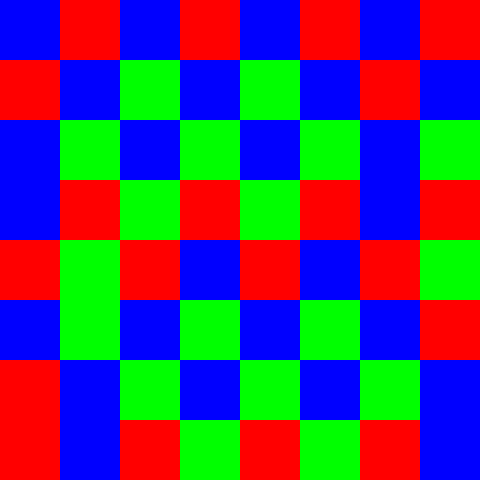}
        \caption{}
    \end{subfigure}
    \caption{Example learned CFAs over training dataset sizes of (a) 250K, (b) 500K, (c) 1 million.}
    \label{fig:scenario_6_cfas}
\end{figure}

One example of learned CFA for each training dataset size is illustrated in Fig.~\ref{fig:scenario_6_cfas}. It can be seen that the learned CFA over the largest training dataset shows more checker-like patterns, more uniformly sampling each color channel. Since each independent training case might result in different CFAs, it is not proper to state that there is only one optimal CFA. Each of our independent training cases resulted in different CFAs and demosaicing models. However, our observation is that the variance on both the learned CFAs as well as the achieved performance in terms of PSNR and SSIM metrics gets lower as training datasets are larger. Due to limited computational resources, we are able to train our models over a maximum of 1 million image patches. However, training the proposed architecture over much larger datasets would be informative on whether the observed trends in performance increases and lower variances continue for larger dataset cases.  

\section{Discussion and Future Work} \label{sec:discussion}
In this paper, we demonstrate a joint architecture both learning a binary color filter array together with demosaicing with a deep neural network. Our results show that the learned CFAs with the proposed architecture result in enhanced reconstruction performance compared to classical fixed CFAs such as Bayer. Since the proposed approach learns to select a single color channel at each pixel, learned CFAs are practical and physically implementable in digital cameras. 

In this section, we would like to discuss the proposed approach and its results in terms of its implications and potential future work. First, the proposed architecture is composed of two submodules: one for learning a binary CFA and the other for reconstructing a full-color image from the sampled CFA output. Both of these modules are learned when we train the architecture jointly. Hence the learned CFA is dependent on the task of the second module, which is demosaicing. Suppose another task such as classification is utilized with a neural network architecture in the second module. In that case, the proposed joint architecture can learn a CFA that would be optimal for that task. Hence the proposed joint architecture allows task-dependent CFA learning which could have other future applications. 

The results from Section~\ref{subsec:comp_filter_size} shows that an optimal CFA size of $8 \times 8$ is observed for the proposed architecture. Any higher filter size resulted in a reduction in reconstruction quality. This could mean that larger and more complex CFA patterns might not be necessary for higher reconstruction quality. An important point is that the any machine learning model depends on the training data and are optimal for that dataset. While this could result in more distinct CFAs for specific applications, it also means a larger training dataset is required to learn a more generalized CFA. The results in Section~\ref{subsec:comp_data_size} show that with increased training dataset sizes even though different CFAs are learned, the average reconstruction performance increases, and the variance in the performance gets smaller. We believe training of the proposed architecture over much larger datasets has the potential to lead to more generalized learned CFAs. 

An interesting observation is the bias toward the blue channel in the learned CFAs. This finding contrasts with the idea of prioritizing the green channel in hand-designed filters as the more informative channel, especially taking into account that green is the least selected color in almost all the learned filters. It is important to include that a few filters end up having more red pixels, but green almost always appears as the least selected pixel. More extensive comparisons with alternative CFA learning methods and better analyses might lead to more definitive answers.

Future work on this study includes analysis on the effect of noise and cross-talk, implementation and analysis of the HardMax module with neural network models for various computational imaging tasks for learning task-specific CFAs, and potential hardware implementation of the proposed filters for a more realistic analysis.

\section{Conclusion} \label{sec:conclusion}
This study presents a binary CFA learning module based on hard thresholding with a deep learning-based demosaicing network in a joint architecture. While a measurement learning approach based on gradient adaptation is developed for binary CFA learning, a demosaicer architecture based on novel DL-based image reconstruction models is jointly learned. The proposed model is trained and tested over Kodak and BSDS500 datasets. Since the proposed approach learns to select a single color channel at each pixel, the learned CFA is easily adaptable to modern commercial cameras. Both RGB and RGBW CFAs can be learned with the proposed approach and increased reconstruction performance in PSNR and SSIM metrics are achieved compared to both fixed well-known filters such as Bayer or alternative learned filters. 

\bibliographystyle{ieeetr}
\bibliography{ref}

\end{document}